%% file: main_aa.tex
\documentclass[longauth]{aa} 

\makeatletter
\@ifpackageloaded{lineno}{%
  \AtBeginDocument{%
    \nolinenumbers
    \let\linenumbers\relax
    \let\endlinenumbers\relax
    \providecommand\internallinenumbers{}%
    \renewcommand\makeLineNumber{\relax}%
  }%
}{}
\makeatother

\usepackage{graphicx}
\usepackage{txfonts}
\usepackage{orcidlink}

\usepackage{hyperref}	
\hypersetup{colorlinks=true, linkcolor=blue, citecolor=blue, filecolor=blue, urlcolor=blue}

\input{commands}

\begin{document} 
\nolinenumbers

\title{The PHANGS-MUSE/HST-H$\alpha$ Nebulae Catalogue}
\subtitle{Parsec-Scale Resolved Structure, Physical Conditions, and Stellar Associations across Nearby Galaxies}

\authorrunning{Barnes et al.}

\input{authours}

\date{Received 30/05/2025; accepted 01/10/2025}
 
\abstract
{
We present the PHANGS-MUSE/HST-H$\alpha$ nebulae catalogue, comprising \Nhst\ spatially resolved nebulae across 19 nearby star-forming galaxies ($D<20$\,Mpc), based on high-resolution H$\alpha$ imaging from HST, homogenised to a fixed (10\,pc) physical resolution and sensitivity. Combined with MUSE integral field spectroscopy, this enables robust classification of \Nhii\ \ion{H}{ii} regions and the separation of planetary nebulae and supernova remnants. We derive electron densities for \Nne\ \ion{H}{ii} regions using [S\,\textsc{ii}] diagnostics and adopt direct or representative electron temperatures for consistent physical characterisation. Nebular sizes are measured using circularised radii and intensity-weighted second moments, yielding a median radius of approximately 20\,pc and extending down to (sub)-parsec (deconvolved) radii. A structural complexity score is introduced via hierarchical segmentation to trace substructure, highlighting that around a third of the regions are \hii\ complexes containing several individual clusters/bubbles, with an increased fraction of these regions in galactic centres. A luminosity–size relation, calibrated using the resolved HST sample, is applied to \Nmuseall\ MUSE nebulae, allowing the recovery of nebular sizes down to $\sim$1\,pc and providing statistical completeness beyond the HST detection limit. Comparisons with classical Strömgren radii indicate that observed sizes are systematically larger, corresponding to typical volume filling factors with a \textcolor{\fontcolor}{median of $\epsilon \sim 0.22$ (10th–90th percentile 0.06–0.78), with larger regions exhibiting progressively lower values.} We associate \Nassoc\ \ion{H}{ii} regions with stellar populations from the PHANGS-HST association catalogue, finding median ages of $\sim$3\,Myr and typical stellar masses of around $10^4$–$10^5$\,M$_\odot$, supporting the link between ionised nebular and young stellar populations. We also assess the impact of diffuse ionised gas on emission-line diagnostics and, after removing confirmed supernova remnants, find no strong variation in line ratios with nebular resolution, indicating minimal systematic bias in the MUSE catalogue. This dataset establishes a detailed, spatially resolved connection between nebular structure and ionising sources, and provides a benchmark for future studies of feedback, DIG contributions, and star formation regulation in the ISM, especially in combination with matched high-resolution observations. \textcolor{\fontcolor}{The full catalogue is made publicly available in machine-readable format.}
}

\maketitle
   
%
\section{Introduction}
\label{sec_int}

\input{figures/fig_maps}

Emission lines from ionised gas play a central role in our understanding of galaxy evolution over cosmic time \citep[see the review by][]{Kewley2019}. These lines enable the determination of spectroscopic redshifts and provide critical diagnostics of key physical properties of galaxies, including star formation rates, the presence of active galactic nuclei (AGN), ionised gas kinematics, gas-phase metallicities, extinction, and more \citep[e.g.][]{Baldwin1981, Kewley2001, Sanchez2024}. Past and ongoing observational surveys have leveraged advances in instrumentation - such as long-slit spectral stepping (e.g. the TYPHOON survey; \citealt{Ho2017, Grasha2022}), imaging Fourier transform spectrographs (e.g. SIGNALS with SITELLE on the CFHT; \citealt{Rousseau-Nepton2019}), and integral field spectroscopy (IFS; e.g. CALIFA, MaNGA, SAMI, and SDSS V LVM \citep{Sanchez2012b, Bundy2015, Bryant2015, Drory2024}, as well as MAD, TIMER, PHANGS, and GECKOS with MUSE/VLT; \citealt{Erroz-Ferrer2019, Gadotti2019, Emsellem2022, vandeSande2024}) - to map multiple emission lines across thousands of galaxies.

However, the individual ionised nebulae responsible for these emission lines, such as \ion{H}{ii} regions, planetary nebulae, supernova remnants, and the narrow and broad-line regions of AGN, are typically much smaller than the $\sim$\,100\,pc spatial resolution achievable in ground-based surveys of galaxies out to only a few Mpc \citep[e.g.][]{Schinnerer2024}.
Fully spatially resolved studies of individual nebulae are therefore largely confined to the nearest systems, where ground-based observations can achieve physical resolutions of $\sim$\,10\,pc or better \citep[e.g.][]{Relano2009, Lopez2011, Lopez2014, McLeod2020, DellaBruna2020, Drory2024, Kreckel2024}. This resolution gap limits the statistical power available to study the physical properties of nebulae across large samples of galaxy environments. In particular, it hinders the ability to determine the true physical extents of nebulae, characterise their internal morphology, disentangle overlapping structures, and distinguish compact sources from the surrounding diffuse ionised gas \citep[DIG; e.g.][]{Reynolds1990, Haffner2009}, introducing systematic uncertainties in efforts to study the physical conditions within nebulae and the impact of stellar feedback on the interstellar medium \citep[ISM; see e.g.][]{Barnes2021, Barnes2022, Pathak2025}. 

In this work, we combine newly obtained H$\alpha$ narrow-band imaging from the PHANGS-HST survey \citep{Chandar2025}, \textcolor{\fontcolor}{with physical resolutions of a few parsecs to 10\,pc (for galaxy distances between 5 and 20\,Mpc), and complementary MUSE integral field spectroscopy from the PHANGS-MUSE survey \citep{Emsellem2022}, with physical resolutions between a few 10s to 100\,pc. In synergy, the HST H$\alpha$ narrow-band imaging enables detailed measurements of the resolved size, shape, and internal structure of each nebula, while the MUSE data provide key (previously unresolved) physical properties such as extinction, density, and ionisation state. Additionally, we incorporate broadband HST observations from the PHANGS-HST stellar association catalogue \citep{Larson2023}, which allow us to identify the embedded stellar populations powering each nebula. This unique combination connects the resolved morphology of the ionised gas with both its physical state and its stellar energy sources, enabling a complete view of the energy and momentum budget driving the evolution of each region.}

\textcolor{\fontcolor}{This work presents a catalogue of over 5000 resolved nebulae across a representative sample of nearby star-forming galaxies --- the largest to date to combine morphological, spectroscopic, and stellar population information at parsec-scale resolution. Quantifying nebular sizes and morphologies provides new constraints on feedback-regulated star formation, including porosity and filling factors (\S\,\ref{subsec_size_strom}), the diffuse ionised background (\S\ref{subsec_BPT_DIG}), and enables tests of feedback-driven expansion (bubble growth, e.g. \citealp{Watkins2023a, Watkins2023b}). The resolved morphologies also allow direct comparison with bubble structures traced by mid-IR PAH emission in JWST observations, linking ionised gas to photodissociation regions at bubble interfaces (e.g. \citealp{Barnes2023}). Looking ahead, the catalogue provides an essential reference point for multi-wavelength studies: JWST (e.g. see right panel of Fig.\,\ref{fig_maps}) and ALMA data will extend these analyses into the embedded and molecular phases of star formation, while radio and near-IR tracers (e.g. Pa$\alpha$, Br$\alpha$; \citealt{Pedrini2024}) provide dust-penetrating views of ionised gas. More broadly, the catalogue establishes a uniform, statistically powerful baseline against which detailed follow-up studies and theoretical models can be compared, offering a starting point for testing star formation efficiency, clustered feedback, and ISM regulation on galactic scales.}

\textcolor{\fontcolor}{This paper is structured as follows. Section~\ref{sec_obs} describes the PHANGS-MUSE sample and the new PHANGS-HST H$\alpha$ observations. Section~\ref{sec_cat} details the construction of the PHANGS-MUSE/HST-H$\alpha$ nebulae catalogue, including source identification, measurement of physical and morphological properties, and association matching. An example of the final catalogue is presented in Appendix~\ref{appedix_catalogue}, where we also provide a description of the delivered data products. The complete catalogue is available online via CDS, together with associated masks and maps, alongside a table documenting all columns. Section~\ref{sec_catprops} presents the resulting population statistics, compares HST- and MUSE-based measurements, and places our results in context with prior studies. Section~\ref{sec_summary} summarises the main findings and outlines future directions.}

\section{Observations}
\label{sec_obs}

\input{figures/fig_maps_zoom_single}

This work primarily leverages observations taken as part of the PHANGS survey, which was designed specifically to resolve galaxies into individual elements of the star-formation process: molecular clouds, \ion{H}{ii} regions, and stellar clusters (see survey papers \citealp{Leroy2021a, Lee2022, Lee2023, Emsellem2022, Chandar2025}; and also Figs.\,\ref{fig_maps} and \ref{fig_maps_zoom_single}).

\subsection{PHANGS-MUSE}
\label{subsec_phangsmuse}

In this work, we focus on the 19 nearby star-forming galaxies spectroscopically mapped as part of the PHANGS--MUSE survey \citep[see][]{Emsellem2022}. 
This sub-sample of PHANGS galaxies spans a broad range of massive ($9.4 < \log M < 10.8$) spiral galaxies that lie along the main sequence of star-forming galaxies.
These MUSE observations have a wavelength coverage of 4800 to 9300 \AA, and achieved an average angular resolution of around 1\arcsec\ (or around 50\,pc at 10\,Mpc). Further details on the sample, observations, reduction, and MUSE data products are provided in \citet{Emsellem2022}. 

\subsection{The PHANGS-MUSE nebular catalogue}

We use the PHANGS-MUSE nebular catalogue \citep{Santoro2022, Groves2023}, with updated auroral line fits (and associated properties) from \citet{Brazzini2024}.\footnote{Also see the catalogues produced using the same PHANGS-MUSE observations but via an independent method by \citet{Congiu2023}.} 
The catalogue inherits the typical $\sim$1 arcsec angular resolution of the MUSE observations (corresponding to $\sim$50\,pc at 10\,Mpc; see \S\,\ref{subsec_phangsmuse}), which sets the effective spatial resolution for all nebular measurements. This resolution enables the detection of thousands of distinct ionised nebulae across the 19 galaxies in the PHANGS-MUSE sample, but with the limitation of still blending compact substructures (see \S\,\ref{subsec_hstha}).

The PHANGS-MUSE nebular catalogue was produced using a modified version of the {\sc HIIphot} algorithm \citep{Thilker2000}, as initially adapted by \citet{Kreckel2019} for application within the PHANGS-MUSE survey.
The {\sc HIIphot} algorithm functions by identifying significant, isolated peaks in the H$\alpha$ emission map. These peaks represent potential nebulae, allowing the algorithm to systematically detect local maxima that are sufficiently spatially separated to be considered distinct structures. This method minimises the inclusion of diffuse emission and overlapping regions, effectively isolating nebulae. For each detected peak, the algorithm also defines spatial boundaries based on the H$\alpha$ intensity contours, segmenting each nebula from the surrounding interstellar medium and nearby structures.

Across the full sample of 19 MUSE galaxies, \citet{Groves2023} identified \Nmuseall\ distinct nebulae as independent peaks in H$\alpha$ emission. Note that the MUSE coverage and HST coverage do not exactly match (see Fig.\,\ref{fig_maps}), such that the total number of nebulae from the MUSE catalogue included within the HST \ha\ coverage (see \S\,\ref{subsec_hstha}) is \Nmuse\ ($\sim$\,80\%; see Tab.\,\ref{tab_map_compprops}). These numbers exclude sources whose footprint falls within the (foreground) star masks, and also nebulae with centres within 1 PSF FWHM of the edges of the PHANGS–MUSE galaxy footprints \citep[see][]{Groves2023}.

A radius is determined for each nebula from the circularised enclosed area, $r_\mathrm{circ} = \sqrt{A/\pi}$, where $A$ is the area from \citet{Groves2023}, calculated using the distances in Tab.\,\ref{tab_galprops}. The integrated MUSE spectra provided flux measurements for key emission lines, including H$\alpha$, H$\beta$, [\ion{O}{iii}], and [\ion{N}{ii}]. The dust extinction was measured using the Balmer decrement, and strong-line prescriptions were applied to derive gas-phase metallicities and ionisation parameters (see \citealp{Groves2023} for details). Moreover, \citet{Groves2023} distinguished between the \hii\ region population and other nebulae (e.g. planetary nebulae or supernova remnants) using the BPT \citep{Baldwin1981} classification criteria from \citet{Kewley2001} and \citet{Kauffmann2003}. There are \Nmusehii\ nebulae classified as \hii\ regions out of the \Nmuse\ nebulae within the catalogue covered by the HST observations. Lastly, \citet{Groves2023} also categorised nebulae based on their galactic environment (centre, bar, arm, inter-arm, and disc) using the environmental masks produced by \citet{Querejeta2021}. 

\subsection{PHANGS-HST-H$\alpha$}
\label{subsec_hstha}

The 19 galaxies in our sample (see Tab.\,\ref{tab_galprops}) were observed with high-angular-resolution H$\alpha$ narrow-band imaging as part of the PHANGS-HST Treasury program (PID 17126; PI Chandar) and a complementary General Observer program (PID 17457; PI Belfiore). Full details of the observations, data reduction, and final products are given in \citet{Chandar2025}; here we provide a brief summary.

For each galaxy, either the F658N or F657N narrow-band filter was used, selected according to the galaxy’s systemic velocity to capture the H$\alpha$ 6563\,\AA\ line. The observations were configured to maximise overlap with the five broad-band filters from the PHANGS-HST Treasury Survey (F275W, F336W, F438W, F555W, and F814W; Cycle 26, PID 15654; PI Lee), ensuring robust continuum subtraction and synergy with the stellar population data. 

Prior to continuum subtraction, both the narrow-band (F658N or F657N) and the adjacent broad-band filters (F555W and F814W) were background-subtracted using synthetic filter images generated from the PHANGS-MUSE observations and corrected for Milky Way foreground extinction. The continuum-subtracted narrow-band images were then corrected for contamination by the [N\,\textsc{ii}]$\lambda6548$ and [N\,\textsc{ii}]$\lambda6583$ emission lines using PHANGS-MUSE spectroscopy.

The final HST H$\alpha$ maps achieve an angular resolution of $\sim$0.1\arcsec\ (corresponding to $\sim$2.5–10\,pc across the sample) with a pixel scale of 0.04\arcsec\ ($\sim$1–4\,pc). This procedure yields high-fidelity emission-line maps for each galaxy, which form the basis of the following analysis (see \citealp{Chandar2025} for further details).

\subsection{PHANGS-HST association catalogues}
\label{subsec_hstassosiations}

In order to investigate the connection between ionised nebulae and their associated stellar populations, we make use of the multi-scale stellar association catalogues produced as part of the PHANGS–HST high-level data products \citep{Larson2023}. These catalogues provide a comprehensive census of stellar populations, encompassing both compact clusters and diffuse associations. This contrasts with the compact cluster catalogues, which primarily trace the densest peaks of star formation and omit much of the lower-density component \citep{Maschmann2024}. 

The identification process begins with point-like sources detected in the HST images using DOLPHOT \citep{Dolphin2016}, based on either NUV- or V-band photometry. A smoothed tracer image is then created by convolving the spatial distribution of sources to a set of predefined scales (typically 8, 16, 32, or 64 pc). Local background subtraction is performed by removing a version of the tracer image smoothed to four times the original scale. The watershed algorithm \citep{vanderWalt2014} is then applied to the background-subtracted tracer image to delineate the boundaries of stellar associations, identified as localised overdensities at each scale. Photometry of the stars within each association provides fluxes across the five HST filters, which are then used to derive ages, masses, and extinctions by fitting theoretical single stellar population models \citep{Bruzual2003} with CIGALE \citep{Boquien2019}, assuming a fully sampled initial mass function \citep{Chabrier2003}. 

Caveats associated with these catalogues — particularly concerning the reliability of the lowest masses and youngest ages — are discussed in Section.~\ref{subsec_stellarprops}.

\section{PHANGS-MUSE/HST-H$\alpha$ nebulae catalogue}
\label{sec_cat}


The PHANGS-MUSE/HST-H$\alpha$ nebulae catalogue is constructed using the PHANGS-MUSE nebulae catalogue as a prior (see Fig.\,\ref{fig_maps_zoom_single}). This “top-down” approach — applying the MUSE nebulae as masks before running source identification on the HST images — is preferred over a “bottom-up” method, where structures are first identified in the HST images and then matched to MUSE nebulae. The top-down method ensures direct consistency with the PHANGS-MUSE catalogue, facilitates the incorporation of MUSE spectroscopic information, and mitigates the risk of incompleteness introduced by the lower surface brightness sensitivity of the HST narrow-band imaging. While the HST observations provide higher spatial resolution (and therefore increased complexity), they are also intrinsically less sensitive to diffuse, low-surface-brightness emission than the MUSE data \citep{Chandar2025}. A similar top-down methodology was adopted by \citet{Barnes2022} to identify and analyse a small sample of \hii\ regions in NGC\,672.


\subsection{Data homogenization}
\label{subsec_DataHomogenization}

This process of identifying and cataloguing the sources within each galaxy was challenging due to the range in spatial scales and complexity of each nebula. This is a result of the factor of $\sim$\,4 range of physical resolution achieved across the galaxy sample, the varying noise profiles within and among individual galaxies (see \citealp{Chandar2025}), as well as the inherent nature of the nebulae themselves.

To mitigate these effects, and to provide a consistent analysis across all galaxies in the sample, we homogenised the PHANGS-HST H$\alpha$ images before performing our source identification (\S\,\ref{subsec_sourceident}). This was a two-step process: 
\begin{enumerate}
    \item \textbf{Resolution normalisation:} All images were convolved to a fixed physical resolution of 10\,pc, assuming a Gaussian point-spread function (PSF). This corresponds to the angular resolution of HST at a distance of 20\,Mpc, and is close to the lowest physical resolution in our sample (NGC\,1365 at 19.6\,Mpc; see Tab.\,\ref{tab_galprops}).
    \item \textbf{Noise normalisation:} For each galaxy, a Gaussian noise map was generated and smoothed using the same Gaussian kernel as applied to the science data. The smoothed noise maps were scaled such that, when added in quadrature to the observed images, the resulting maps achieved a uniform noise level with standard deviation $\sigma_\mathrm{final,10pc} = 6.25 \times 10^{-16}$\,erg/s/cm$^2$/arcsec$^{2}$ (see Appendix\,\ref{appendix_data_homogenisation} for additional details and caveats). This value was chosen as the maximum noise level measured across the 10\,pc-smoothed images. 
\end{enumerate}
The procedure above produced maps with the same physical resolution and noise levels, which we will use for the following source identification. 

\subsection{Source identification}
\label{subsec_sourceident}

\input{tables/tab_map_compprops}

We begin by applying the PHANGS-MUSE nebula masks to the homogenised PHANGS-HST H$\alpha$ images. Adopting a top-down approach, each of the \Nmuse\ MUSE nebulae located within the overlapping coverage is analysed individually. A series of methods for source identification in the HST H$\alpha$ images were tested; we ultimately adopted a threshold-based masking procedure, which offered a straightforward and reproducible means of isolating emission features across the full dynamic range and diverse morphologies in the sample.

This approach follows the two-threshold Boolean masking \textcolor{\fontcolor}{technique originally developed for molecular cloud identification in spectral line data cubes \citep{Rosolowsky2006} and subsequently implemented for the PHANGS–ALMA survey \citep{Leroy2021a, Rosolowsky2021}.} Adapting this framework to our HST narrow-band imaging, we use a high threshold of $5\sigma_\mathrm{final,10pc}$ and a low threshold of $2\sigma_\mathrm{final,10pc}$ (see Sect.~\ref{subsec_DataHomogenization}). For each masked MUSE nebula, binary masks are first generated at both thresholds. Structures smaller than three times the smoothed PSF area are removed from the low-threshold mask, since regions smaller than this are unlikely to yield reliable physical measurements, while the high-threshold mask is pruned of sources smaller than one PSF area to exclude spurious compact peaks. The cleaned high-threshold mask is then grown into the low-threshold mask, ensuring that compact, high-surface-brightness emission is retained while simultaneously recovering fainter diffuse substructure. This procedure also suppresses isolated noise fluctuations that would otherwise bias size and flux estimates. Finally, any internal holes are filled by including all pixels within the outermost contiguous boundary, so that each nebula mask corresponds to a single, continuous region.


Potential contaminants within the catalogue were assessed through a combination of parameter-based criteria and manual inspection. Firstly, flags were assigned to regions intersecting the edges of the HST coverage, following the convention established in the MUSE catalogue \citep{Groves2023}. An additional flag was used to identify regions in contact with neighbouring sources, which may form part of a larger complex. Secondly, a detailed manual inspection was conducted to identify residual contaminants. These typically manifest as small noise peaks, artefacts, or sources exhibiting luminosities significantly higher than expected relative to MUSE. A total of 576 regions were examined, selected based on either a physical size smaller than 10\,pc, fewer than 20 pixels, or luminosities exceeding the MUSE measurements, specifically where $L_\mathrm{H\alpha}(\mathrm{HST}) / L_\mathrm{H\alpha}(\mathrm{MUSE}) > 1$. Among these, we identified 30 bright foreground stars not effectively flagged in the MUSE catalogue, typically characterised by high-luminosity point sources, strong counterparts in the HST broadband images, and prominent negative artefacts in the continuum-subtracted \ha\ images. In addition, 54 cosmic rays or image artefacts were identified, appearing as compact, bright point sources, often located near the edges of the map. A single background galaxy was also flagged in NGC\,1087. All identifications have been incorporated into an additional dedicated manual check flag within the catalogue.


The final catalogue comprises \Nhstnoflags\ nebulae in the PHANGS-HST-H$\alpha$ maps across the galaxy sample, corresponding to approximately one fifth of the total number of nebulae identified in the PHANGS-MUSE catalogue within the matched area (\Nmuse). 
Incorporating the HST edge flags, MUSE edge flags, and manual contamination flags described above, we obtain a total of \Nhst\ regions. This set constitutes the primary science sample used for all statistical analyses and figures in this work, unless otherwise stated.

\subsection{Nebula properties, sizes and luminosities}
\label{subsec_sourceprops}

The physical properties of each nebula identified in the PHANGS-HST-H$\alpha$ images are compiled into a unified catalogue. Each entry in the PHANGS-MUSE/HST-H$\alpha$ nebula catalogue corresponds directly to a unique region in the PHANGS-MUSE nebula catalogue. The final combined catalogue thus contains the full set of nebular parameters from the MUSE-based catalogue described in \citet{Groves2023}, including updated auroral-line fits and derived properties from \citet{Brazzini2024}, as well as the additional parameters derived from the HST H$\alpha$ data in this work (e.g. size and luminosity).

The physical size of each nebula is estimated using two complementary definitions. The first is the circularised radius, defined as $r_\mathrm{circ} = \sqrt{A/\pi}$, where $A$ is the projected area enclosed by the source mask. This definition is consistent with that adopted for the MUSE catalogue. The second is the second-moment radius, $r_\mathrm{mom}$, computed as the geometric mean of the intensity-weighted second spatial moments of the emission within the source boundary (see Sect.~\ref{subsec_size} for a comparative analysis of $r_\mathrm{circ}$ and $r_\mathrm{mom}$).

The total H$\alpha$ flux of each region is calculated by summing the flux within all pixels enclosed by the final source mask. This flux is compared to the corresponding MUSE-based flux measurements in Appendix.~\ref{subsec_luminosity}. To estimate extinction-corrected H$\alpha$ fluxes and luminosities, we correct for the per-region dust-extinction in the PHANGS-MUSE catalogue. The corresponding H$\alpha$ luminosity is computed as $L_{\Ha} = 4 \pi D^2 F_{\Ha}$, where $D$ is the distance to the galaxy and $F_{\Ha}$ is the extinction-corrected flux.

\subsection{Complexity score}
\label{subsec_complexity}

A limitation of our source identification method is its inability to separate multiple sources within a single PHANGS-MUSE catalogue mask (e.g. Figs.~\ref{fig_maps} and \ref{fig_maps_zoom_single}). This leads to source confusion in regions of high structural complexity, where overlapping nebular substructures are not individually resolved (see also Fig.~\ref{fig_maps_zoom_complex}).

To provide a quantitative measure of this internal substructure, we perform a hierarchical segmentation analysis using the \textsc{astrodendro} software \citep{Rosolowsky2008}, applied to the flux distribution within the final HST catalogue mask of each region. The segmentation parameters were chosen to match those adopted in our initial source identification (see Sect.~\ref{subsec_sourceprops}): the minimum number of pixels per structure is set to include one 10\,pc resolution element (\textsc{min\_npix}); the minimum flux threshold is set to \textsc{min\_value} = $2\sigma_\mathrm{final,10pc}$; and the minimum isocontour separation is \textsc{min\_delta} = $5\sigma_\mathrm{final,10pc}$.

\textcolor{\fontcolor}{For each nebula, we define a complexity score ($\mathcal{C}$) as the total number of dendrogram structures identified at all hierarchical levels. This metric is intended primarily as a relative proxy for structural richness rather than an absolute measurement. While the exact value of $\mathcal{C}$ varies with the adopted thresholds, tests varying the parameters by factors of two confirm that the relative ranking of regions is preserved: nebulae classified as complex remain distinct from those with simple morphologies. In this sense, $\mathcal{C}$ provides a practical first-order indicator of relative complexity. The resulting scores range from 0 to $\sim$100 and are grouped into three broad categories:}


\begin{itemize}
\setlength\itemsep{1em}
    \item \Csimple: A score of $\mathcal{C} \leq 1$ (\Nhstsimple\ regions) identifies nebulae with relatively simple morphology. These typically consist of a single weak and diffuse structure ($\mathcal{C} = 0$), or a single compact or brighter diffuse component ($\mathcal{C} = 1$).
    \item \Cintermediate: Regions with $2 \leq \mathcal{C} \leq 5$ (\Nhstintermediate\ regions) exhibit modest substructure, such as multiple compact emission peaks or a combination of diffuse and clumpy features.
    \item \Ccomplex: A score of $\mathcal{C} > 5$ (\Nhstcomplex\ regions) corresponds to morphologically complex nebulae, containing numerous substructures.
\end{itemize}

\textcolor{\fontcolor}{
The classification thresholds were set empirically to best reflect the observed morphologies. Representative examples are shown for NGC\,1566 in Fig.~\ref{fig_maps_zoom_single} (lower left), alongside the corresponding HST H$\alpha$ emission (upper right), with further examples in Fig.~\ref{fig_maps_zoom_complex}. These illustrate that the adopted categories capture the broad contrast between simple, intermediate, and complex nebulae. To preserve flexibility, the individual $\mathcal{C}$ values are reported in the catalogue, so that alternative thresholds can be adopted if required by future analyses. Table~\ref{tab_map_compprops} lists the number of regions in each category per galaxy.
}

\subsection{\ion{H}{ii} region properties}
\label{subsec_hiiregionprops}

We find that \Nhii\ nebulae, corresponding to approximately 94\% of the total \Nhst\ sample, are classified as \hii\ regions using the emission-line diagnostics outlined in \citet{Groves2023}. For this subset, we derive additional physical properties, including electron densities and ionising photon production rates.

Electron densities, $n_\mathrm{e}$, are determined using the \textsc{PyNeb} package \citep{Luridiana2015},\footnote{\textsc{PyNeb} version `1.1.24', which includes atomic information for \ion{S}{ii} taken from \citet{Tayal2010} and \citet{Rynkun2019}.} which solves the equilibrium level populations for user-defined ionic species. For each \hii\ region, $n_\mathrm{e}$ is computed from the flux ratio \Rsii\ $= F_{\SIIa} / F_{\SIIb}$, combined with an assumed electron temperature, $T_\mathrm{e}$. We adopt measured values of $T_\mathrm{e}$ from \citet{Brazzini2024}, derived from the Nitrogen auroral line ratio, where available. In total, \NTe\ \hii\ regions (14\% of the sample) have such measurements. For the remaining majority, we assume a representative value of $T_\mathrm{e} = 8000$\,K, which corresponds to the mean of the measured values (standard deviation is 1800\,K).\footnote{The electron temperature is computed from the auroral-to-nebular line ratio $(\nii\lambda6584 + \nii\lambda6548) / \nii\lambda5755$, using \textsc{PyNeb} with an assumed density of 100\,cm$^{-3}$. The derived $T_\mathrm{e}$ values are insensitive to this assumption at the typical densities of our sample ($<1000$\,cm$^{-3}$; see \citealt{Mendez-Delgado2023}).}

The \Rsii\ diagnostic saturates in the low- and high-density regimes, approaching values of $\sim$\,1.45 for $n_\mathrm{e} \lesssim$ a few 10\,cm$^{-3}$, and $\sim$\,0.4 for $n_\mathrm{e} \gtrsim$ a few $10^3$\,cm$^{-3}$. Within this range, \Rsii\ provides a robust estimate of the [S\,\textsc{ii}]-weighted mean electron density of the ionised gas. Following \citet{Barnes2021}, we exclude regions where \Rsii\ lies within $3\sigma$ of the low-density limit, to avoid unreliable estimates.\footnote{We correct the [\ion{S}{ii}] flux uncertainties from the \citet{Groves2023} catalogue following \citet{Barnes2021}, with a temperature-dependent lower limit estimated using \textsc{PyNeb}. We find a slightly larger correction factor of 1.53, compared to the original value of 1.38.} After applying these criteria, we obtain reliable $n_\mathrm{e}$ estimates for \Nne\ \hii\ regions -- approximately half of the sample. This is comparable to the number of regions with $n_\mathrm{e}$ measurements in the full MUSE sample \citep{Barnes2021}, indicating that the vast majority of denser sources are recovered here.\footnote{Note that a slightly different version of the PHANGS-MUSE catalogue is used in this work.}

We also estimate the ionising photon production rate, $Q$, for each region under the assumption of Case B recombination and optically-thick conditions. For an electron temperature of $T_\mathrm{e} = 8000$\,K and $n_\mathrm{e} < 10^6$\,cm$^{-3}$, $Q \approx L_{\Ha} / (0.45\,h\nu_{\Ha})$, where $L_{\Ha}$ is the extinction-corrected H$\alpha$ luminosity \textcolor{\fontcolor}{(derived from the HST H$\alpha$ flux)}, $h$ is Planck’s constant, and $\nu_{\Ha}$ is the frequency of the H$\alpha$ transition. The factor 0.45 represents the fraction of total hydrogen recombinations that produce H$\alpha$ emission at this temperature (see \citealp{Storey1995, Osterbrock2006, Byler2017}, also Tab.\,14.2 in \citealp{Draine2011b}).

\subsection{Stellar properties}
\label{subsec_stellarprops}

To connect the physical properties of the ionised gas within \hii\ regions (see \S\,\ref{subsec_sourceprops}) to their underlying stellar populations, we link each region to the PHANGS-HST stellar associations catalogue (see \S\,\ref{subsec_hstassosiations}). An example is shown in of Fig.\,\ref{fig_maps_zoom_single} (lower left), with additional examples illustrated in Figs.\,\ref{fig_maps_zoom_complex}. This procedure follows the method introduced by \citet{Scheuermann2023}, who matched stellar associations from HST to the PHANGS-MUSE nebulae catalogue.

We construct a matched catalogue by identifying cases where the HST nebula mask of a given \hii\ region spatially overlaps with one or more stellar associations. A match is accepted if any part of the association boundary lies within the boundary of the PHANGS-MUSE/HST-H$\alpha$ nebula mask. If multiple associations are found within a single nebula mask,\footnote{For example, in the sample of 32\,pc-scale NUV-selected associations, 25\% of matched \hii\ regions are associated with more than one stellar association.} we flag the region accordingly and adopt the properties of the \textit{youngest} stellar association, under the assumption that it is most likely to be responsible for the observed ionisation.

The matching includes associations identified at all spatial scales (8, 16, 32, and 64\,pc) and in both the NUV- and V-band multi-scale association catalogues. However, in the analysis that follows, we focus on the 32\,pc-scale associations derived from the NUV images. As shown by \citet{Scheuermann2023}, the NUV selection better traces young, massive stars, while the 32\,pc scale provides a good match to the resolution of the MUSE data. A total of \Nassoc\ \hii\ regions (64\% of the sample) are matched with at least one NUV-selected 32\,pc-scale stellar association. A per-galaxy breakdown is provided in Table~\ref{tab_map_compprops}.

Table~\ref{tab_agemass_assoc} summarises the 10th, 50th (median), and 90th percentiles of the age and stellar mass (log scale) distributions of the matched associations across the sample. The stellar mass distribution is dominated by associations in the range $\log(M_\ast/\mathrm{M}_\odot) = 3.5$--$4.5$, although several galaxies contain more massive associations (with 90th percentile $\log(M_\ast/\mathrm{M}_\odot) > 5$). These higher-mass associations are typically found in galaxies with elevated star formation rates. For example, NGC\,1365 - the most actively star-forming galaxy in our sample - hosts the most massive associations in terms of median stellar mass. 

We find that the median ages vary modestly between galaxies, reflecting differences in recent star formation history and evolutionary stage, but the overall distribution is strongly peaked toward young associations, with a global median age of 3\,Myr (see Fig.\,\ref{fig_hist_assoc} for the full range of ages). This supports the expected interpretation that most \hii\ regions are powered by the youngest stellar associations in each system.

Before proceeding, we note \textcolor{\fontcolor}{three} important caveats. Firstly, the age distribution of associations exhibits a pronounced peak at 1\,Myr (Fig.\,\ref{fig_hist_assoc}), which arises from an artefact of the SED-fitting methodology used in the PHANGS-HST catalogue \citep{Larson2023}. This bias, inherent to models excluding nebular emission \citep{Bruzual2003}, leads to an over-representation of 1\,Myr ages, as discussed by \citet{Thilker2025}. Consequently, associations assigned an age of 1\,Myr should be interpreted as spanning 1--3\,Myr. Future catalogue versions will incorporate improved models that include nebular emission \citep{Thilker2025}, resolving this degeneracy (see also \citealp{Whitmore2025, Henny2025}). Secondly, the assignment of a single age per \hii\ region or complex does not capture the intrinsic age spread within these regions. Sequential star formation, often driven by feedback mechanisms \citep[e.g.,][]{Elmegreen1977,Walborn1992,Walborn1997,Whitmore2010,Whitmore2025}, leads to a range of stellar ages, typically spanning 1--10\,Myr. This age spread should be considered when interpreting analyses that assume coeval stellar populations. \textcolor{\fontcolor}{Thirdly, the stellar masses reported for associations with $\log(M_\ast/M_\odot) \lesssim 4$ should be regarded with some caution. At these low masses, stochastic sampling of the stellar initial mass function can introduce significant scatter in the derived properties, since the assumption of a fully populated \citet{Chabrier2003} IMF is less appropriate. While such associations are retained in the catalogue for completeness, their mass estimates carry high uncertainties.}

\section{Nebulae catalogue properties}
\label{sec_catprops}

In this section, we provide an overview of key properties derived from the PHANGS-MUSE/HST-H$\alpha$ nebulae catalogue.

\subsection{Structure and completeness}
\label{subsec_structure}

\input{figures/fig_completeness}

As illustrated in Fig.~\ref{fig_maps_zoom_single} (see also Fig.\,\ref{fig_maps_zoom_complex}), the order-of-magnitude improvement in angular resolution provided by HST relative to MUSE enables a significantly more detailed view of the internal structure of the nebulae. In this example, taken from NGC\,1566 \citep{Chandar2025}, the HST data reveal compact emission peaks and substructures that are unresolved in the MUSE observations. Moreover, the HST emission appears more spatially confined, predominantly tracing the higher surface brightness components, while the MUSE data capture more extended, diffuse emission due to their higher surface brightness sensitivity.

In Fig.~\ref{fig_completeness}, we present the completeness of the HST catalogue as a function of MUSE H$\alpha$ luminosity, measured both in terms of the number of recovered regions and their cumulative luminosity. Completeness is computed per luminosity bin by comparing the subset of MUSE nebulae that are matched to HST detections with the total number or luminosity of MUSE nebulae in that bin. At the peak of the MUSE luminosity function (log\,$L_\mathrm{H\alpha}$\,$\sim$\,37), the completeness in number is approximately 5\%, and in luminosity, approximately 10\%. The 50\% completeness thresholds in both number and luminosity are reached at log\,$L_\mathrm{H\alpha}$\,$\sim$\,38. For brighter regions (log\,$L_\mathrm{H\alpha}$\,$>$\,38.5), the completeness rises to 90\% in number and 75\% in luminosity.

These results confirm that the PHANGS-MUSE/HST-H$\alpha$ catalogue is significantly more complete for higher-luminosity nebulae in the MUSE sample, with reduced sensitivity to diffuse, low-luminosity regions due to the limitations of our HST narrow-band imaging.

\subsection{Size}
\label{subsec_size} 

\input{figures/fig_histrad}
\input{figures/fig_histrad_ratio}
\input{figures/fig_histradratioenv}
\input{figures/fig_hist_literature}
\input{figures/fig_histradmoms}

\subsubsection{Comparison of HST and MUSE sizes}

In Fig.~\ref{fig_histrad}, we present the distribution of circularised radii measured from both the PHANGS-MUSE and PHANGS-HST-H$\alpha$ observations across the full galaxy sample. We find that the radii of sources in the MUSE catalogue are clustered around 70\,pc, consistent with the average linear resolution limit of the MUSE observations for this galaxy sample \citep[see also][]{Barnes2021, Santoro2022, Groves2023}, and as listed in Table~\ref{tab_galprops}. By contrast, the HST-based radius distribution is broader, with a peak around 20\,pc and values extending down to the fixed resolution limit of $\sim$\,10\,pc.

The ratio of HST to MUSE circularised radii, $r_\mathrm{circ}(\mathrm{HST})/r_\mathrm{circ}(\mathrm{MUSE})$, is shown as a grey histogram in Fig.~\ref{fig_histrad_ratio}. The distribution peaks at a ratio of approximately 0.3 and declines smoothly between $\sim$0.1 and 1.0. By construction, the HST-defined nebulae are constrained to lie within the corresponding MUSE masks, and therefore this ratio cannot exceed unity.

Also shown in Fig.~\ref{fig_histrad_ratio} (coloured histograms) is the distribution of size ratios subdivided by complexity score class (\Csimple, \Cintermediate, \Ccomplex; see Sect.~\ref{subsec_complexity}). Regions with high size ratios ($r_\mathrm{HST}/r_\mathrm{MUSE} > 0.6$) are predominantly found among the complex class (\Ccomplex), whereas regions with low size ratios ($< 0.4$) are dominated by simple nebulae (\Csimple). This indicates that large HST–MUSE size ratios typically arise from spatially blended or confused regions in the MUSE data — corresponding to complex, extended emission — while smaller ratios reflect more compact, isolated nebulae well resolved by HST. Given the correlation between size and H$\alpha$ luminosity (see Fig.~\ref{fig_lumrad_fitting}), this also implies that more luminous nebulae tend to be more complex.

To assess how these ratios vary across different galactic environments, we show in Fig.~\ref{fig_histradenv} the distribution of HST/MUSE size ratios as a function of environment classification, following the definitions in \citet{Querejeta2021}. Regions located in galaxy centres show the highest fraction (approximately 25\%) of near-unity size ratios, consistent with expectations of crowding and structural complexity, as well as higher and more complex background, diffuse emission, in the central regions (see also Fig.~\ref{fig_maps}).\footnote{The segmentation used to construct the PHANGS--MUSE catalogue is known to struggle in very crowded circumnuclear environments, such as starburst rings (e.g. NGC\,1300, NGC\,1512, NGC\,1672, NGC\,3351, NGC\,4303, NGC\,4321) and in galaxies hosting low-luminosity AGN (e.g. NGC\,1365, NGC\,1512, NGC\,1566, NGC\,1672; see \citealt{Santoro2022, Groves2023}). In these environments, deblending individual \ion{H}{ii} regions is not possible at the MUSE resolution, which then contributes to the larger and more complex nebulae we recover in the central regions. These regions are nonetheless included in the catalogue presented here, but their derived sizes should be interpreted with caution.} In contrast, the bar, arm, inter-arm, and disc environments exhibit lower fractions of high-ratio regions. Each of these environments contains at least 50\% of its regions with HST/MUSE size ratios below 0.4.

A Kruskal–Wallis test \citep{Kruskal1952} confirms significant differences across the five environments ($H=432.3$, $p<0.001$). Pairwise Kolmogorov–Smirnov tests show that the centre differs significantly from all other regions (KS = 0.54--0.64; $p<0.001$). While the non-central environments are broadly similar, significant differences exist between most pairs (KS $\sim$0.1; $p<0.05$), except for the inter-arm and disc, which do not differ significantly (KS = 0.048; $p=0.27$).


\subsubsection{Comparison of HST sizes to other surveys}

We also compare the size distribution derived from the PHANGS-MUSE/HST-H$\alpha$ catalogue with that obtained in higher spatial resolution extragalactic and Galactic studies. As shown in Fig.~\ref{fig_hist_lit}, the radius distribution of our sample closely matches that of NGC\,300 observed with MUSE (at a distance of around 2\,Mpc, the achieved resolution of $\sim$\,1\arcsec\ matches our homogenised spatial resolution of 10 pc; \citealp{McLeod2021}. This agreement reflects the similarly high spatial resolution achieved in both datasets, enabling detection of compact \hii\ regions down to scales of a few parsecs. We note that this comparison is most appropriate for the subset of simple regions in our catalogue (\Csimple), which are expected to best represent isolated, well-resolved nebulae in external galaxies.

In contrast, the distribution of Galactic \hii\ region sizes from \citet{Anderson2014} peaks at smaller values. This is attributable to the higher linear resolution (at sub-pc for the closest regions, and ) achievable within the Milky Way, as well as to differences in selection methodology (i.e. using different emission -- dust and PAH emission -- to identify their regions). In particular, their catalogue is constructed from 22\,$\mu$m emission and likely includes a large number of very young, embedded \hii\ regions that are optically obscured and intrinsically smaller in physical size.

The PHANGS-MUSE/HST-H$\alpha$ sample includes a broader distribution of sizes, extending to significantly larger radii due to the presence of more complex and blended regions, particularly in crowded environments. Nevertheless, the overlap in the size distributions with both Galactic and extragalactic studies supports the reliability of our catalogue in capturing a representative range of nebular sizes, particularly among the more luminous and spatially extended \hii\ region population.

\subsubsection{Comparison of HST sizes ($r_\mathrm{circ}$ vs $r_\mathrm{mom}$ vs $r_\mathrm{mom, deconv}$)}

A potential limitation of using the circularised radius, $r_\mathrm{circ}$, as a size metric is its sensitivity to both the intrinsic emission morphology of the region and the local noise properties of the observations. In particular, extended low-surface-brightness wings or fragmented structures can artificially increase the enclosed area, leading to overestimated sizes. To mitigate this, we also consider an alternative size definition based on the intensity-weighted second spatial moment of the H$\alpha$ emission, denoted $r_\mathrm{mom}$. This moment-based size is further deconvolved with the 10\,pc FWHM point-spread function, assuming a Gaussian profile, following the relation $r_\mathrm{mom,deconv} = \sqrt{r_\mathrm{mom}^{2} - \sigma_\mathrm{PSF}^{2}}$  where $\sigma_\mathrm{PSF} = \mathrm{FWHM} / \sqrt{8 \ln 2}$ corresponds to the standard deviation of the (assumed to be Gaussian) point-spread function. 

In Fig.~\ref{fig_histradmoms}, we compare the distributions of $r_\mathrm{mom}$, $r_\mathrm{mom,deconv}$ and $r_\mathrm{circ}$ across the PHANGS-MUSE/HST-H$\alpha$ catalogue. The $r_\mathrm{mom}$ and $r_\mathrm{mom,deconv}$ distributions are systematically shifted toward smaller radii, with a peak near 10\,pc, while $r_\mathrm{circ}$ peaks around 20–25\,pc. The distribution of the $r_\mathrm{mom}/r_\mathrm{circ}$ ratio peaks just below a value of 0.5, similar to the ratio between the standard deviation of a Gaussian profile and its half-width at tenth maximum (HWTM), for which $\sigma/\mathrm{HWTM} \approx 0.47$.

This then reflects the different physical interpretations of each metric: $r_\mathrm{mom}$ (and $r_\mathrm{mom,deconv}$) is most analogous to the standard deviation of the surface brightness distribution, providing a compactness-weighted size, whereas $r_\mathrm{circ}$ reflects the total projected area above a threshold, more sensitive to irregular shapes and diffuse substructure. The difference between these two estimators is therefore a useful diagnostic of nebular morphology and may serve as a proxy for structural concentration.

\section{Analysis}

\input{figures/fig_lumrad}

\subsection{Recovering sizes for unresolved regions}
\label{subsec_radlumscaling}

Although a substantial number of \hii\ regions are recovered in our HST-based catalogue (\Nhst), a large fraction of those detected by MUSE remain undetected. 
Specifically, $\sim$80\% of the MUSE nebulae within the matched field of view are not identified in the HST data. 
Hence, to estimate the size distribution for this large population, we adopt the methodology introduced by \citet{Pathak2025},\footnote{The results presented here differ slightly from those of \citet{Pathak2025}, who used an earlier internal release of the HST catalogue at native resolution and noise levels.} which involves calibrating a luminosity–size relation using MUSE-derived H$\alpha$ luminosities and HST-based size measurements.

Figure~\ref{fig_lumrad_fitting} shows the resulting luminosity–size distributions for various combinations of luminosity and size measurements from MUSE and HST. These distributions span approximately two orders of magnitude in radius and four orders of magnitude in luminosity. We see tight correlation using resolved size estimates from the HST catalogue comparing to either the luminosity measured from the HST or MUSE. 
We fit a linear function in log–log space to both the median-binned and all the points within a 1 to 99 percentile range (see Figure~\ref{fig_lumrad_fitting}). The fits are given as the following: \\\\
\textbf{Point fits:}
\begin{align}
 \log(r_\mathrm{circ}) & = 0.479 \log(L_\mathrm{H\alpha}) - 16.820, \\
 \log(r_\mathrm{mom})  & = 0.467 \log(L_\mathrm{H\alpha}) - 16.678, \\
 \log(r_\mathrm{mom,deconv}) & = 0.523 \log(L_\mathrm{H\alpha}) - 18.859. \label{eq_lumradfit_mom_points}
\end{align}
\textbf{Bin fits:}
\begin{align}
 \log(r_\mathrm{circ}) & = 0.451 \log(L_\mathrm{H\alpha}) - 15.742, \\
 \log(r_\mathrm{mom})  & = 0.428 \log(L_\mathrm{H\alpha}) - 15.194, \\
 \log(r_\mathrm{mom,deconv}) & = 0.470 \log(L_\mathrm{H\alpha}) - 16.824. \label{eq_lumradfit_mom_bin}
\end{align}
We use these relations to scale the MUSE-derived H$\alpha$ luminosities and recover corresponding radius estimates, which we denote $r_\mathrm{lum, circ}$, $r_\mathrm{lum, mom}$, or $r_\mathrm{lum, mom, deconv}$. 

This approach enables us to estimate sizes for the full sample of MUSE-detected nebulae, circumventing the completeness limits of the HST catalogue. Notably, the recovered distributions extend to significantly smaller physical scales, reaching down to $\sim$1\,pc in the case of $r_\mathrm{lum, mom}$. This reflects the inclusion of lower-luminosity regions from the MUSE catalogue, which, under the assumed scaling relation, correspond to compact nebulae that would not be detected in the HST imaging.


\textcolor{\fontcolor}{We note that this method implicitly assumes a universal luminosity–size relation across all \hii\ regions. While this is a strong assumption, it is physically motivated by the expected connection between stellar mass, ionising photon output, and nebular luminosity (e.g. \citealt{Brown2021}). We demonstrate this in Section\,\ref{subsec_stellarprops_comp} (Fig.\,\ref{fig_rad_age_mass_assoc}), where we show that \ha\ luminosity correlates strongly with the stellar masses of associated populations, supporting the use of this scaling relation for statistical recovery of sizes.} However, if the regions undetected in the \textit{HST} imaging are preferentially diffuse and spatially extended, then their sizes would be systematically underestimated. In this case, the use of a luminosity–size relation calibrated on higher surface brightness regions would map their MUSE-derived luminosities onto radii that are too small, primarily biasing the low-luminosity tail of the recovered size distribution and potentially distorting interpretations of the population’s structural properties.


\subsection{Systematic comparison to Strömgren sphere models}
\label{subsec_size_strom}

\input{figures/fig_rad_strom}
\input{figures/fig_fillingfraction}

Here, we assess whether the observed sizes of \hii\ regions are consistent with predictions from idealised Strömgren sphere models \citep{Stromgren1939}. The expected radius of a Strömgren sphere is given by
\begin{equation}
\label{eq_strom}
r_\mathrm{str} = \left(\frac{3Q}{4\pi \alpha_B(T_\mathrm{e}) n_\mathrm{e}^2} \right)^{1/3},
\end{equation}
where $Q$ is the ionising photon flux, $n_\mathrm{e}$ is the hydrogen number density, and $\alpha_B(T_\mathrm{e})$ is the temperature-dependent case B recombination coefficient (in units of cm$^{3}$,s$^{-1}$). For $\alpha_B(T_\mathrm{e})$, we adopt the fitting formula from \citet{Hui1997}, based on \citet{Ferland1992}:
\begin{equation}
\label{eq_alpha}
\alpha_\mathrm{B}(\Te) = 2.753\times10^{-14} \left(\frac{\Te}{315614}\right)^{-1.5} \left[1 + \left(\frac{115188}{\Te}\right)^{0.4}\right]^{-2.2},
\end{equation}
where \Te\ is the electron temperature in Kelvin. For regions lacking direct temperature estimates from auroral lines (e.g.\ $\nii\lambda5755$), we adopt a representative value of 8000\,K (see Sect.\ref{subsec_sourceprops}).

Figure~\ref{fig_rad_strom} compares the measured sizes of \hii\ regions to the theoretical Strömgren radii computed from Eq.~\ref{eq_strom}. We find that $r_\mathrm{str}$ is systematically smaller than all observed size estimates, with typical offsets ranging from factors of a few (compared to $r_\mathrm{mom, deconv}$) up to an order of magnitude (relative to $r_\mathrm{circ}(\mathrm{MUSE})$).

A natural interpretation of this discrepancy is that the ionised gas traced by [S\textsc{ii}] emission occupies only a small fraction of the total nebular volume. Because the Strömgren radius scales as $r_\mathrm{str} \propto n_\mathrm{e}^{-2/3}$, electron densities inferred from [S\textsc{ii}] are biased high in the presence of clumping or edge-brightened shells, which enhance the [S\textsc{ii}] emissivity and thereby reduce the inferred $r_\mathrm{str}$. This behaviour is expected from theoretical models and seen in resolved observations, where D-type ionisation fronts sweep up dense shells around expanding \hii\ regions, and is consistent with the edge-brightened morphologies seen in our sample (e.g. see \citealp{Weilbacher2015, Mcleod2019, Kreckel2024} for resolved observation examples). 

\textcolor{\fontcolor}{
To quantify this effect, we estimate effective volume filling factors by comparing the predicted Strömgren radii to observed sizes (as introduced by \citealp{Osterbrock1959}). The filling factor can be written as,
\begin{equation}
\epsilon = \frac{3Q}{4\pi\,\alpha_B(T_\mathrm{e})\,n_\mathrm{e}^2\,r^3} \;=\; \frac{r_\mathrm{str}^3}{r^3},
\label{equ_fillingfactor}
\end{equation}
where $Q$ is the ionising photon rate, $n_\mathrm{e}$ the [S\textsc{ii}]-derived electron density, and $r$ the observed radius. This expression is equivalent to the formulation used in ionisation-parameter studies of \hii\ regions \citep[e.g.][]{Diaz1991}. In Fig.\,\ref{fig_rad_filling}, we show $\epsilon$ as a function of the moment-based radius ($r_\mathrm{mom}$), which provides an upper-limit estimate of the filling factor. We find that compact regions on $\sim$10\,pc scales approach $\epsilon \sim 1$ (completely filled), while progressively larger regions exhibit systematically lower values. Across the full sample we obtain a median $\epsilon = 0.22$ (mean 0.40), with a 10th--90th percentile range of 0.06--0.78. These values are consistent with independent determinations for nearby \hii\ regions: \citet{Kennicutt1984} reported typical filling factors of 0.01--0.1, while \citet{Cedres2013} found values in the range $10^{-6}$--$10^{-1}$ with a similar decreasing trend of lower filling factors for larger regions. This reinforces the conclusion that [S\textsc{ii}] preferentially traces dense, low-filling-factor structures embedded within larger ionised volumes.}


\subsection{Comparison to stellar properties}
\label{subsec_stellarprops_comp}

\input{figures/fig_histassociations_agemass}
\input{figures/fig_rad_mass}

As shown in Fig.\,\ref{fig_hist_assoc}, the stellar mass and age distributions of the PHANGS--HST association catalogue (see Sect.~\ref{subsec_hstassosiations} and \ref{subsec_stellarprops}) vary systematically with the adopted association scale. \textcolor{\fontcolor}{Note that here the ``scale'' refers to the Gaussian-smoothing lengths used in the multi-scale watershed segmentation of \citep{Larson2023}, and should not be confused with the physical radii of the \hii\ regions measured in this work.} The stellar mass distributions are skewed toward lower-mass systems across all identification scales, with the median association mass decreasing from the 64\,pc to the 8\,pc smoothing scales. At the same time, the age distributions exhibit a secondary peak around 7\,Myr at the largest (64\,pc) scale, reflecting the fact that larger apertures encompass composite associations that blend multiple stellar populations of different ages and masses. By contrast, the smaller apertures preferentially isolate individual, younger, and lower-mass associations. This behaviour is consistent with the hierarchical nature of star formation, in which stellar structures are organised over a range of spatial scales, from sub-cluster groupings to larger associations and complexes \citep[e.g.][]{Grasha2017, Larson2023}.

In Fig.~\ref{fig_rad_age_mass_assoc}, we explore the relationship between the H$\alpha$-defined \hii\ region luminosity and size, and the mass of the associated stellar population (here using the 32\,pc-scale, NUV-selected associations). A clear positive correlation is observed: more massive associations tend to power larger and more luminous ionised regions. This is consistent with theoretical expectations, where more massive clusters generate stronger ionising radiation fields, resulting in larger ionised volumes \citep[e.g.][]{Lopez2014}, as predicted by classical Strömgren sphere models \citep{Stromgren1939}.

In contrast, the left-hand panel of Fig.~\ref{fig_rad_age_mass_assoc} shows little correlation between region size and association age. This suggests that, within our sample, the primary driver of nebular size is the mass of the powering stellar association, rather than its age. This interpretation is consistent with the rapid decline in UV luminosity beyond a few Myr \citep[e.g.][]{Leitherer1999}, and with the relatively narrow age spread of the associated clusters.


\subsection{Does DIG systematically bias line ratios in unresolved \ion{H}{ii} regions?}
\label{subsec_BPT_DIG}

\input{figures/fig_bpt}

We assess whether diffuse ionised gas (DIG) significantly biases the nebular line ratios measured in unresolved \ion{H}{ii} regions in MUSE by examining correlations between emission-line diagnostics and spatial resolution, as traced by the radius ratio between HST and MUSE detections. Figure~\ref{fig_bpt} presents the variation of several commonly used diagnostic ratios — [O\textsc{iii}]/H$\beta$, [N\textsc{ii}]/H$\alpha$, [S\textsc{ii}]/H$\alpha$, [O\textsc{i}]/H$\alpha$ — along with H$\alpha$ equivalent width, as a function of the HST-to-MUSE radius ratio ($r_{\mathrm{HST}}/r_{\mathrm{MUSE}}$).

To ensure that our analysis is not biased by shock-dominated sources, we have removed all supernova remnants (SNRs) and SNR candidates identified in the MUSE nebular catalogue by \citet{Li2024}. In total, \Nsnr\ (10$\%$ of our sample) such regions are excluded from this analysis, as SNRs typically exhibit elevated low-ionisation line ratios that would skew the trends under investigation.

In general, the emission-line ratios exhibit only weak or negligible dependence on the radius ratio (see Fig.\,\ref{fig_bpt}). The exception is the H$\alpha$ equivalent width, which increases steadily with radius ratio, likely reflecting the contribution of more extended, actively star-forming complexes in regions that are better resolved in the MUSE observations. Among the diagnostic line ratios, a mild enhancement is seen at small radius ratios ($r_{\mathrm{HST}}/r_{\mathrm{MUSE}} \lesssim 0.2$), possibly indicating an increased DIG contribution. However, the lowest-radius-ratio bin, where these deviations are most pronounced, contains relatively few sources, limiting its statistical significance.

To quantify the global impact, we find that from $r_{\mathrm{HST}}/r_{\mathrm{MUSE}} = 0.2$ to 0.8, the median binned values of [O\textsc{iii}]/H$\beta$, [N\textsc{ii}]/H$\alpha$, [S\textsc{ii}]/H$\alpha$, and [O\textsc{i}]/H$\alpha$ differ from the global sample median by less than 25\%. This suggests that DIG contamination does not systematically bias the diagnostic line ratios for the majority of sources in the MUSE nebular catalogue. We therefore conclude that, although DIG may locally affect the emission-line properties of compact, unresolved sources, its overall impact on the catalogue-wide line ratio distributions is limited. These findings support the reliability of the MUSE \ion{H}{ii} region catalogue, with minimal systematic bias introduced by unresolved DIG contamination.

\section{Summary and outlook}
\label{sec_summary}

We present the PHANGS-MUSE/HST-H$\alpha$ nebulae catalogue, comprising \Nhst\ spatially resolved ionised nebulae across 19 nearby galaxies ($D<20$\,Mpc), based on high-resolution HST narrow-band imaging homogenised to a uniform physical resolution and sensitivity. Combined with PHANGS-MUSE spectroscopy, this dataset enables robust classification of \Nhii\ \ion{H}{ii} regions, as well as the identification and separation of planetary nebulae and supernova remnants.

Key physical properties, including sizes, electron densities, and ionising photon production rates, are derived for each region. A median \ion{H}{ii} region radius of $\sim$20\,pc is measured, with sizes extending down to the 10\,pc resolution limit. We introduce a complexity score to quantify internal nebular substructure, revealing that approximately one-third of regions comprise morphologically complex \ion{H}{ii} complexes containing multiple clusters and bubbles, particularly in galaxy centres.

A luminosity–size relation is calibrated using the HST-resolved regions and applied to the full PHANGS-MUSE nebular catalogue, enabling the recovery of physical sizes down to $\sim$1\,pc and correcting for incompleteness in the HST catalogue. Observed nebular sizes systematically exceed those predicted by classical Strömgren sphere models, corresponding to typical volume filling factors with a median of $\epsilon \sim 0.22$ (10th–90th percentile 0.06–0.78). Compact $\sim$10\,pc regions approach unity, while progressively larger regions exhibit systematically lower filling factors, consistent with the clumpy, shell-dominated morphologies seen in resolved nebulae.

By associating nebulae with stellar populations from the PHANGS-HST catalogue, we find that most \ion{H}{ii} regions are powered by young (median age $\sim$3\,Myr) stellar populations with typical masses of $10^4$--$10^5$\,M$_\odot$. A positive correlation between region size and stellar mass is observed, consistent with expectations from photoionisation models. We find no significant systematic biases in nebular emission-line ratios due to unresolved diffuse ionised gas, supporting the reliability of MUSE spectroscopy for global diagnostic studies.

The PHANGS-MUSE/HST-H$\alpha$ catalogue provides a comprehensive, spatially resolved view of ionised nebulae across a representative sample of nearby galaxies. This dataset enables detailed studies of nebular structure, feedback processes, and the connection between star formation and the interstellar medium. In future work, we aim to combine this catalogue with high-resolution data from JWST, ALMA, and other facilities to further characterise the multi-phase interstellar medium and the role of stellar feedback in shaping galaxy evolution.

The complete catalogue, including all \Nhstnoflags\ nebulae and \Ncols\ measured columns, is made publicly available via CDS,  together with a comprehensive README describing all columns. Associated products include region masks (with values corresponding to the \texttt{region\_id} in the catalogue and matched to the PHANGS-MUSE catalogue of \citealt{Groves2023}) and the homogenised HST H$\alpha$ maps convolved to 10\,pc resolution that form the basis of this catalogue.

\begin{acknowledgements}

ATB dedicates this paper to his wife, Christina Barnes, for beginning this new chapter as parents together, and warmly acknowledges their son, Theodore Michael Barnes, for graciously waiting two hours after its submission before beginning his entrance into the world.

The authors of this paper are grateful to the anonymous referee for their constructive and detailed suggestions, which helped significantly improve the quality of this paper.

This work has been carried out as part of the PHANGS collaboration. Based on observations from the PHANGS-MUSE program, collected at the European Southern Observatory under ESO programmes 094.C-0623 (PI: Kreckel), 095.C-0473,  098.C-0484 (PI: Blanc), 1100.B-0651 (PHANGS-MUSE; PI: Schinnerer), as well as 094.B-0321 (MAGNUM; PI: Marconi), 099.B-0242, 0100.B-0116, 098.B-0551 (MAD; PI: Carollo) and 097.B-0640 (TIMER; PI: Gadotti).

In addition, this research is based on observations made with the NASA/ESA Hubble Space Telescope obtained from the Space Telescope Science Institute, which is operated by the Association of Universities for Research in Astronomy, Inc., under NASA contract NAS 5–26555. These observations are associated with programs 15654, 17126, and 17457. 

FB acknowledges support from the INAF Fundamental Astrophysics program 2022.
KK gratefully acknowledges funding from the Deutsche Forschungsgemeinschaft (DFG, German Research Foundation) in the form of an Emmy Noether Research Group (grant number KR4598/2-1, PI Kreckel) and the European Research Council’s starting grant ERC StG-101077573 (“ISM-METALS").
SCOG and RSK acknowledge financial support from the European Research Council via the ERC Synergy Grant ``ECOGAL'' (project ID 855130) and from the German Excellence Strategy via the Heidelberg Cluster of Excellence (EXC 2181 - 390900948) ``STRUCTURES''.
KG is supported by the Australian Research Council through the Discovery Early Career Researcher Award (DECRA) Fellowship (project number DE220100766) funded by the Australian Government. OE acknowledges funding from the Deutsche Forschungsgemeinschaft (DFG, German Research Foundation) -- project-ID 541068876.
RSK also acknowledges support from the German Ministry for Economic Affairs and Climate Action in project ``MAINN'' (funding ID 50OO2206). In addition, RSK thanks the 2024/25 Class of Radcliffe Fellows for highly interesting and stimulating discussions. 
ZB gratefully acknowledges the Collaborative Research Center 1601 (SFB 1601 sub-project B3) funded by the Deutsche Forschungsgemeinschaft (DFG, German Research Foundation) – 500700252.
MB acknowledges support from the ANID BASAL project FB210003. This work was supported by the French government through the France 2030 investment plan managed by the National Research Agency (ANR), as part of the Initiative of Excellence of Université Côte d’Azur under reference number ANR-15-IDEX-01.
\end{acknowledgements}

\bibliographystyle{aa}
\bibliography{references, references_extra}

\begin{appendix}

\section{Nebula Catalogue}
\label{appedix_catalogue}

\textcolor{\fontcolor}{The PHANGS-HST H$\alpha$ Nebulae Catalogue (DR1) provides a homogeneous set of measurements for \Nhstnoflags\ regions  (i.e. without any flag-based filtering) identified across the 19 nearby galaxies in the PHANGS-HST sample. Each catalogue entry corresponds to a single nebula, with positions, sizes, fluxes, luminosities, and a variety of derived physical parameters. These include electron densities, ionisation parameters, and metallicities from matched PHANGS-MUSE spectroscopy, as well as stellar association ages and masses from the PHANGS-HST multi-scale stellar association 
catalogues.}

\textcolor{\fontcolor}{Table~\ref{tab_ned_cat} illustrates a representative subset of the catalogue, showing a few rows and a selection of columns. The full catalogue contains \Ncols\ columns, spanning structural, photometric, spectroscopic, and environmental properties. A complete description of all parameters included in the table is provided in Table~\ref{tab_ned_cat_desc1} and \ref{tab_ned_cat_desc2}. The catalogue itself, including the full sample of \Nhstnoflags\ regions, together with the complete documentation, is accessible online via CDS.}

\input{tables/table_catalogue}
\input{tables/table_catalogue_descriptions}

\section{Galaxy Sample}

Table~\ref{tab_galprops} provides an overview of the properties of the 19 galaxies included in this study.

\input{tables/tab_galprops}

\section{Maps}

Figure\,\ref{fig_maps_zoom_complex} show additional example nebulae exhibiting simple, intermediate, and complex morphologies. 

\input{figures/fig_maps_zoom}




\section{Luminosity distributions and flux recovery}
\label{subsec_luminosity}

\input{figures/fig_histlum}
\input{figures/fig_radlumratios}

In Fig.~\ref{fig_histlum}, we present the H$\alpha$ luminosity distributions for all regions in the PHANGS-MUSE/HST-H$\alpha$ nebula catalogue. When comparing the subset of sources matched between the HST and MUSE catalogues, we find that both distributions peak at similar luminosities in the range $L_\mathrm{H\alpha} \sim 10^{37}$–$10^{38}$\,erg\,s$^{-1}$, with a systematic offset in the median luminosity of approximately a factor of two (see lower panel of Fig.~\ref{fig_histlum}).\footnote{This comparison includes only regions that are identified in both the HST and MUSE catalogues. For the full luminosity distribution of all MUSE regions, see Fig.~\ref{fig_completeness}.}

Given that the HST observations are both lower in sensitivity and can detect more compact structures than MUSE, we do not expect the HST catalogue to recover the full H$\alpha$ flux for most regions. Nonetheless, we find that a small fraction (48) of regions exhibit $L_\mathrm{H\alpha}(\mathrm{HST})$ significantly (50\%) higher than $L_\mathrm{H\alpha}(\mathrm{MUSE})$ (see inset in Fig.~\ref{fig_histlum}), which, are limited to known problematic regions (e.g. centres or particularly complex regions).

In Fig.~\ref{fig_radlumratios}, we show the ratio $r_\mathrm{circ}(\mathrm{HST})/r_\mathrm{circ}(\mathrm{MUSE})$ as a function of $r_\mathrm{circ}(\mathrm{HST})$, $r_\mathrm{circ}(\mathrm{MUSE})$, $L_\mathrm{H\alpha}(\mathrm{HST})$, and $L_\mathrm{H\alpha}(\mathrm{HST})/L_\mathrm{H\alpha}(\mathrm{MUSE})$. As expected, larger HST regions are more comparable in size to their MUSE counterparts. Moreover, we find that the larger and more luminous nebulae are those for which a higher fraction of the total MUSE flux is recovered in the HST data. The final panel shows that regions with more similar sizes between HST and MUSE also exhibit more similar H$\alpha$ luminosities, indicating that spatially extended regions are better matched across both catalogues.

\section{Size-Luminosity distributions}

\input{figures/fig_histlumenv}

Figure~\ref{fig_histlumenv} presents the distribution of H$\alpha$ luminosity ratios between HST and MUSE detections, separated by galactic environment (i.e. centre, bar, arm, interarm, and disc regions). The histograms and cumulative distributions indicate that MUSE captures a larger fraction of the total luminosity in central environment, while arm, bar and interarm regions show comparatively lower recovery fractions. This trend reflects underlying environmental differences in gas density, feedback, and star formation structure. These results are consistent with the result that larger, more complex regions (which tend to reside in centres) recover more total flux.

The cumulative distributions confirm that HST observations capture a significant fraction of the H$\alpha$ emission, particularly from smaller and denser nebulae missed in lower-resolution datasets.


\section{Size distribution as function of environment}

In Fig.\,\ref{fig_histradenv_appendix}, we show the size distributions as a function of environment. 

\input{figures/fig_histradenv}







\section{BPT ratios as a function of luminosity}

As discussed in Sect.~\ref{subsec_BPT_DIG}, we assess the impact of diffuse ionised gas (DIG) on nebular line ratios. In Fig.~\ref{fig_bpt_lum}, we reproduce the panels from Fig.~\ref{fig_bpt} but substitute the x-axis with the ratio of the H$\alpha$ luminosities measured with HST and MUSE. This provides an alternative view of how DIG-related dilution may influence diagnostic line ratios.

\input{figures/fig_bpt_lum}

\section{Nebulae and Stellar properties}

Table~\ref{tab_radlum} summarises the statistics of nebular radii and luminosities for each galaxy in the catalogue. 
Table~\ref{tab_agemass_assoc} provides the corresponding stellar properties for matched associations, including their masses and ages.

\input{tables/tab_radlumstats}
\input{tables/tab_agemass_assoc}



\section{Data Homogenisation}
\label{appendix_data_homogenisation}

\input{figures/fig_noise}

\textcolor{\fontcolor}{In order to compare nebular properties across the PHANGS sample, all HST \ha\ images were homogenised to a common physical resolution of 10\,pc by convolution with a Gaussian kernel. The applied noise contribution is computed as 
\begin{equation}
\sigma_\mathrm{applied,10pc}^2 = \sigma_\mathrm{final,10pc}^2 - \sigma_\mathrm{obs,10pc}^2,
\end{equation}
where $\sigma_\mathrm{obs,10pc}$ is the median noise level measured in the 10\,pc-smoothed image.}

\textcolor{\fontcolor}{Several caveats should be noted (see \citealp{Lee2022, Chandar2025} for more details on the specific data reduction for PHANGS-HST). First, the noise properties of HST data products differ from the Gaussian assumption adopted here. Photon-counting statistics yield an intrinsically Poissonian noise distribution, and the drizzling process used to combine dithered exposures introduces correlated noise between neighbouring pixels. As shown by \citet{Fruchter2002}, the correlation arises because input pixels are resampled onto the output grid with fractional overlaps, so that the pixel-to-pixel RMS underestimates the true noise on larger spatial scales.}

\textcolor{\fontcolor}{Second, the HST PSF is not well described by a Gaussian, exhibiting diffraction features and Airy rings. In addition, the PSF varies across the detector field of view, meaning that the effective shape is not spatially invariant. The drizzling process used to combine dithered exposures further modifies the PSF, with the exact outcome depending on the chosen kernel and pixel fraction parameters, typically broadening the core and altering the wings.}

\textcolor{\fontcolor}{Finally, the images contain a mix of correlated and uncorrelated components due to both the detector sampling and the drizzling resampling. While a Gaussian kernel provides a convenient first-order approximation for homogenisation, it cannot perfectly capture these subtleties. Nonetheless, adopting a common Gaussian PSF of 10\,pc allows for consistent comparison across the sample, and we emphasise that the small deviations introduced by this simplification do not affect our primary conclusions. Future work may incorporate more sophisticated homogenisation schemes that explicitly propagate correlated noise and non-Gaussian PSF structure.}

\textcolor{\fontcolor}{Fig.\,\ref{fig_noise} illustrates the impact of smoothing and noise re-scaling for two representative galaxies: NGC\,1300 (upper panels) and NGC\,5068 (lower panels). NGC\,1300, one of the most distant galaxies in the sample (18.99\,Mpc; \citealp{Anand2021a, Anand2021b}), has relatively low intrinsic noise and therefore requires little smoothing but a significant amount of additional noise injection. In contrast, NGC\,5068 is among the nearest galaxies (5.20\,Mpc; \citealp{Anand2021a, Anand2021b}), requiring substantial smoothing and added noise. For each galaxy we show the noise distributions of the native image, the image smoothed to 10\,pc resolution, and the smoothed image with added noise. Gaussian fits to the distributions are overplotted to highlight the similarities and differences. The native images show clear deviations from Gaussian behaviour, with extended tails towards negative values as expected for Poisson statistics (noting that the positive tail may also include contamination from real emission). The smoothed images also deviate from purely Gaussian profiles, though in the case of NGC\,5068 the smoothing strongly reduces the effective noise level (but not for NGC\,1300). Finally, adding noise back in produces images with noise properties more consistent across the sample. While this procedure is not ideal for the reasons discussed above, it provides a practical and uniform approach, which can be improved upon in future work.}

\end{appendix}
\end{document}

%% file: commands.tex
\newcommand{\HII}{\ion{H}{II}}
\newcommand{\hii}{\ion{H}{II}}
\newcommand{\Ha}{H{$\alpha$}}

\def \msun {{\rm M$_\odot$}}

\def \Te {\ensuremath{T_\mathrm{e}}}
\def \ne {\ensuremath{n_\mathrm{e}}}

\def \Ha {\textup{H\ensuremath{\alpha}}}

\newcommand{\SIIa}{[\ion{S}{II}]\ensuremath{\lambda6716}}
\newcommand{\SIIb}{[\ion{S}{II}]\ensuremath{\lambda6731}}

\newcommand{\nii}{[N\,\textsc{ii}]}
\newcommand{\sii}{[S\,\textsc{ii}]}

\newcommand{\ha}{\mbox{$\mathrm{H}\alpha$}}

\newcommand{\Rsii}{\ensuremath{R_{\sii}}}

\newcommand{\Csimple}{$\mathcal{C}_{\mathrm{simp}}$}
\newcommand{\Cintermediate}{$\mathcal{C}_{\mathrm{inter}}$}
\newcommand{\Ccomplex}{$\mathcal{C}_{\mathrm{comp}}$}

\def \Nmuseall {30,790} 
\def \Nmuse {25,910} 
\def \Nmusehii {19,528}
\def \Nhst {5177}
\def \Nhstnoflags {5467}

\def \Nhstsimple  {3386}
\def \Nhstintermediate {1169}
\def \Nhstcomplex {622}

\def \Nhii {4882}
\def \Nne {2544}
\def \Nassoc {3349}

\def \NTe {779}

\def \Nsnr {556}

\def \Ncols {241}

\newcommand{\fontcolor}{black}

%% file: authours.tex
\author{
        A.~T.~Barnes \inst{\ref{eso}}\orcidlink{0000-0003-0410-4504} \and 
        R.~Chandar \inst{\ref{toledo},}\orcidlink{0000-0003-0085-4623} \and  
        K.~Kreckel \inst{\ref{ari}}\orcidlink{0000-0001-6551-3091} \and
        F.~Belfiore \inst{\ref{infa}}\orcidlink{0000-0002-2545-5752} \and 
        D.~Pathak \inst{\ref{ohio}, \ref{columbus}}\orcidlink{0000-0003-2721-487X} \and 
        D.~Thilker \inst{\ref{jhu}}\orcidlink{0000-0002-8528-7340} \and
        A.~K.~Leroy \inst{\ref{ohio}, \ref{columbus}}\orcidlink{0000-0002-2545-1700} \and
        B.~Groves \inst{\ref{uniwestaus}}\orcidlink{0000-0002-9768-0246} \and 
        S.~C.~O.~Glover \inst{\ref{ita}}\orcidlink{0000-0001-6708-1317} \and
        R.~McClain \inst{\ref{ohio}, \ref{columbus}}\orcidlink{0000-0002-6187-4866} \and 
        A.~Amiri \inst{\ref{UOA}} \and
        Z.~Bazzi \inst{\ref{ubonn}}\orcidlink{0009-0001-1221-0975} \and
        M.~Boquien \inst{\ref{unica}}\orcidlink{0000-0003-0946-6176} \and
        E.~Congiu \inst{\ref{esochile}}\orcidlink{0000-0002-8549-4083} \and
        D.~A.~Dale \inst{\ref{wyo}}\orcidlink{0000-0002-5782-9093} \and
        O.~V.~Egorov \inst{\ref{ari}}\orcidlink{0000-0002-4755-118X} \and
        E.~Emsellem \inst{\ref{eso},\ref{cral}}\orcidlink{0000-0002-6155-7166} \and
        K.~Grasha \inst{\ref{anu}}\orcidlink{0000-0002-3247-5321} \and
        J.~Gonzalez Lobos \inst{\ref{mpia}}\orcidlink{0000-0002-6056-3425} \and
        K.~Henny \inst{\ref{wyo}}\orcidlink{0000-0001-7448-1749} \and
        H.~He \inst{\ref{ubonn}}\orcidlink{0000-0001-9020-1858} \and
        R.~Indebetouw \inst{\ref{uva}, \ref{nrao}}\orcidlink{0000-0002-4663-6827} \and
        J.~C.~Lee \inst{\ref{STScI}, \ref{stew}, \ref{Gemini}}\orcidlink{0000-0002-2278-9407} \and
        J.~Li \inst{\ref{ari}}\orcidlink{0000-0002-4825-9367} \and
        F.-H.~Liang \inst{\ref{ari}}\orcidlink{0000-0003-2496-1247} \and
        K.~Larson \inst{\ref{STScI}}\orcidlink{0000-0003-3917-6460} \and
        D.~Maschmann \inst{\ref{stew}}\orcidlink{0000-0001-6038-9511} \and
        S.~E.~Meidt \inst{\ref{gent}}\orcidlink{0000-0002-6118-4048} \and
        J.~Eduardo Méndez-Delgado \inst{\ref{unam}}\orcidlink{0000-0002-6972-6411} \and
        J.~Neumann \inst{\ref{mpia}}\orcidlink{0000-0002-3289-8914} \and
        H.-A.~Pan \inst{\ref{tku}}\orcidlink{0000-0002-1370-6964} \and
        M.~Querejeta \inst{\ref{oan}}\orcidlink{0000-0002-0472-1011} \and
        E.~Rosolowsky \inst{\ref{alberta}}\orcidlink{0000-0002-5204-2259} \and
        S. K. Sarbadhicary \inst{\ref{jhu}}\orcidlink{0000-0002-4781-7291} \and
        F.~Scheuermann \inst{\ref{ari}}\orcidlink{0000-0003-2707-4678} \and
        L.~\'Ubeda \inst{\ref{STScI}}\orcidlink{0000-0001-7130-2880} \and
        T.~G.~Williams \inst{\ref{Ox}}\orcidlink{0000-0002-0012-2142} \and
        T.~D.~Weinbeck \inst{\ref{wyo}}\orcidlink{0009-0005-8923-558X} \and
        B.~Whitmore \inst{\ref{STScI}}\orcidlink{0000-0002-3784-7032} \and
        A.~Wofford \inst{\ref{iaunam}}\orcidlink{0000-0001-8289-3428}
        \and
        the PHANGS collaboration}
\institute{
\label{eso} European Southern Observatory (ESO), Karl-Schwarzschild-Stra{\ss}e 2, 85748 Garching, Germany  \and
\label{toledo} Ritter Astrophysical Research Center, University of Toledo, Toledo, OH, 43606 \and
\label{jhu} Department of Physics and Astronomy, The Johns Hopkins University, Baltimore, MD 21218, USA \and
\label{ari} Astronomisches Rechen-Institut, Zentrum für Astronomie der Universität Heidelberg, Mönchhofstraße 12-14, 69120 Heidelberg, Germany \and 
\label{infa} INAF — Osservatorio Astrofisico di Arcetri, Largo E. Fermi 5, I-50125, Florence, Italy \and
\label{uniwestaus} International Centre for Radio Astronomy Research, University of Western Australia, 7 Fairway, Crawley, 6009 WA, Australia \and
\label{ita} Universit\"{a}t Heidelberg, Zentrum f\"{u}r Astronomie, Institut f\"{u}r Theoretische Astrophysik, Albert-Ueberle-Str.\ 2, 69120 Heidelberg, Germany \and
\label{ohio} Department of Astronomy, Ohio State University, 180 W. 18th Ave, Columbus, Ohio 43210 \and
\label{columbus} Center for Cosmology and Astroparticle Physics, 191 West Woodruff Avenue, Columbus, OH 43210, USA \and
\label{STScI} Space Telescope Science Institute, 3700 San Martin Drive, Baltimore, MD 21218, USA \and
\label{anu} Research School of Astronomy and Astrophysics, Australian National University, Canberra, ACT 2611, Australia \and
\label{Ox} Sub-department of Astrophysics, Department of Physics, University of Oxford, Keble Road, Oxford OX1 3RH, UK \and
\label{wyo} Department of Physics \& Astronomy, University of Wyoming, Laramie, WY 82071, USA \and
\label{esochile} European Southern Observatory (ESO), Alonso de Córdova 3107, Casilla 19, Santiago 19001, Chile \and
\label{unam} Instituto de Astronom\'{\i}a, Universidad Nacional Aut\'onoma de M\'exico, Ap. 70-264, 04510 CDMX, Mexico \and
\label{Radcliffe} {Elizabeth S.\ and Richard M.\ Cashin Fellow at the Radcliffe Institute for Advanced Studies at Harvard University, 10 Garden Street, Cambridge, MA 02138, USA} \and
\label{IWR} {Universit\"{a}t Heidelberg, Interdisziplin\"{a}res Zentrum f\"{u}r Wissenschaftliches Rechnen, Im Neuenheimer Feld 205, D-69120 Heidelberg, Germany} \and
\label{CfA} {Harvard-Smithsonian Center for Astrophysics, 60 Garden Street, Cambridge, MA 02138, USA} \and
\label{oan} {Observatorio Astron{\'o}mico Nacional (IGN), C/ Alfonso XII 3, E-28014 Madrid, Spain} \and
\label{gent}{Sterrenkundig Observatorium, Universiteit Gent, Krijgslaan 281 S9, B-9000 Gent, Belgium} \and
\label{ubonn}{Argelander-Institut f\"ur Astronomie, University of Bonn, Auf dem H\"ugel 71, 53121 Bonn, Germany} \and
\label{iaunam}{Instituto de Astronom\'ia, Universidad Nacional Aut\'onoma de M\'exico, Unidad Acad\'emica en Ensenada, Km 103 Carr. Tijuana-Ensenada, Ensenada, BC, M\'exico}
\label{cral}{} \and
\label{unica}{Université Côte d'Azur, Observatoire de la Côte d'Azur, CNRS, Laboratoire Lagrange, F-06000 Nice, France} \and
\label{uva}{Astronomy Department, University of Virginia,  P.O. Box 400325, Charlottesville, VA, 22904} \and
\label{nrao}{National Radio Astronomy Observatory, 520 Edgemont Rd, Charlottesville, VA, 22903} \and 
\label{UOA}{Department of Physics, University of Arkansas, 226 Physics Building, 825 West Dickson Street, Fayetteville, AR 72701, USA} \and 
\label{mpia}{Max-Planck-Institut f\"{u}r Astronomie, K\"{o}nigstuhl 17, D-69117 Heidelberg, Germany} \and 
\label{tku}{Department of Physics, Tamkang University, No.151, Yingzhuan Road, Tamsui District, New Taipei City 251301, Taiwan} \and 
\label{stew}{Steward Observatory, University of Arizona, Tucson, AZ 85721, USA} \and 
\label{Gemini}{Gemini Observatory/NSF’s NOIRLab, 950 N. Cherry Avenue, Tucson, AZ, 85719, USA} \and
\label{alberta}{Dept. of Physics, 4-183 CCIS, University of Alberta, Edmonton, AB, T6G 2E1, Canada}
}

%% file: figures/fig_maps.tex
\begin{figure*}
    \centering
	\includegraphics[width=\textwidth]{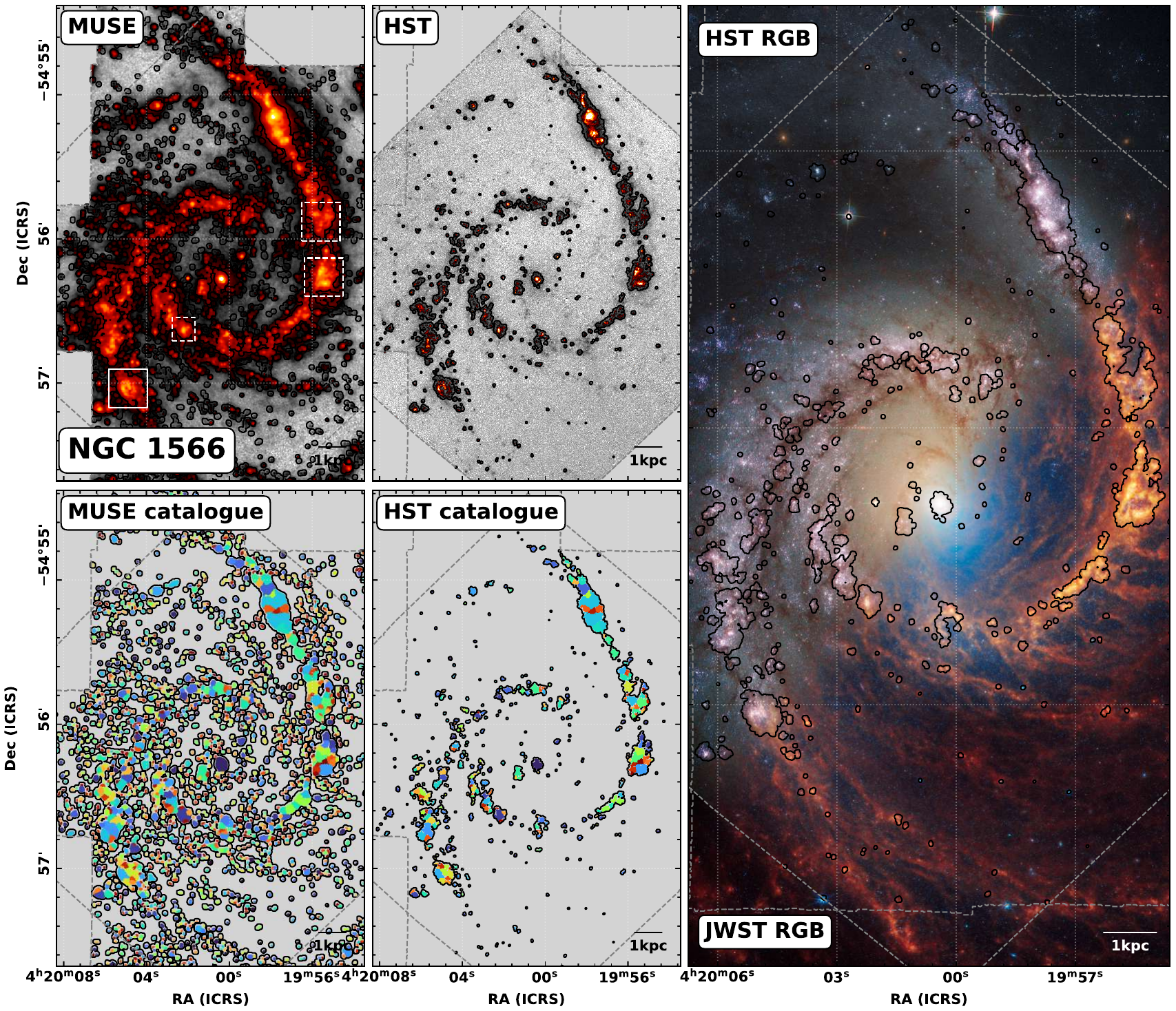}
    \caption{\textbf{Overview of the nebula catalogues towards NGC\,1566.} (\textit{Upper left}) The background colour scale shows the H$\alpha$ emission from the PHANGS-MUSE observations \citep{Emsellem2022}, overlaid with contours showing the boundaries of the sources identified in the PHANGS-MUSE Nebula Catalogue \citep[][see also \citealp{Santoro2022}]{Groves2023}. White boxes indicate the positions of the regions shown in Figs.\,\ref{fig_maps_zoom_single} (solid lines) and \,\ref{fig_maps_zoom_complex} (dashed lines). (\textit{Upper centre}) The background colour scale shows the H$\alpha$ emission from the PHANGS-HST H$\alpha$ observations \citep{Chandar2025}, overlaid with contours showing the boundaries of the sources identified in this work in the PHANGS-HST Nebula Catalogue. (\textit{Bottom left}) The PHANGS-MUSE Nebula Catalogue masks and (\textit{bottom centre}) PHANGS-HST Nebula Catalogue masks are shown with the same colour scale, where the colours denote the region ID. (\textit{Right}) Composite of separate exposures acquired with JWST using the NIRCam and MIRI instruments \citep{Lee2023, Williams2024}, and the HST using the ACS/WFC instrument \citep{Lee2022}. For the JWST part of the image, the assigned colors are Red = F2100W + F1130W + F1000W + F770W, Green = F770W + F360M, Blue = F335M + F300M. For the HST part of the image, the assigned colors are Red = F814W + F656N, Green = F555W, Blue = F435W. Overlaid as contours is the PHANGS-HST Nebula Catalogue (as in the other panels).}
    \label{fig_maps}
\end{figure*}

%% file: figures/fig_maps_zoom_single.tex
\begin{figure*}[!t]
    \centering
        \includegraphics[width=\textwidth]{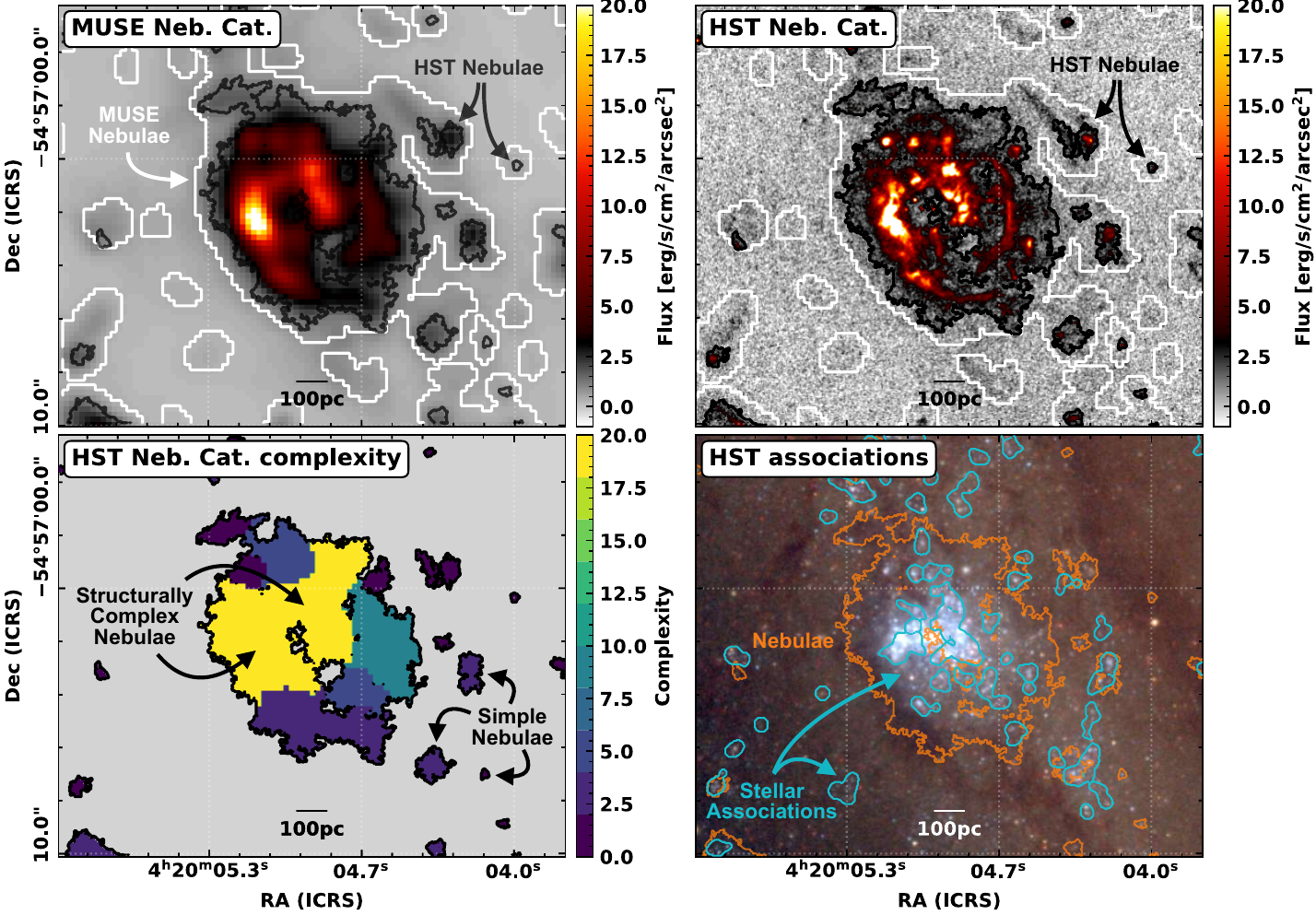} \\ \vspace{-1mm}
    \caption{\textbf{Example of a \hii\ complex region in NGC\,1566 (see \S\,\ref{subsec_complexity}).} (\textit{upper left panel}) MUSE observations and (\textit{upper right}) HST \Ha\ observations that are smoothed to a physical scale of 10\,pc and with a fixed noise level (see \S\,\ref{subsec_DataHomogenization}). Each of these are overlaid with contours showing the boundary of the sources identified in the MUSE (white) and HST (black) observations. (\textit{lower left panel}) A map of the complexity score for the region (see \S\,\ref{subsec_structure}). (\textit{lower right panel}) We show the nebula (orange contours) and 32\,pc NUV-identified  association (\citealp{Larson2023}; cyan contour) overlaid on a HST filter red (F814W) green (F555W) blue (F438W+F336W) image (see \citealp{Lee2022}). See Fig.\,\ref{fig_maps_zoom_complex} for additional examples of simple, intermediate and complex nebulae within NGC\,1566.}
    \label{fig_maps_zoom_single}
\end{figure*}

%% file: tables/tab_map_compprops.tex
\begin{table*}
\centering
    \caption{Completeness statistics for number of nebula the MUSE and HST catalogues.}
\label{tab_map_compprops}

\begin{tabular}{lcccccccccc}
\hline\hline 
    Galaxy & $N_\mathrm{MUSE,tot}$ & $N_\mathrm{MUSE}$ & $N$ & $f_{N}$ & $N_{\mathcal{C}_\mathrm{simp}}$ & $N_{\mathcal{C}_\mathrm{inter}}$ & $N_{\mathcal{C}_\mathrm{comp}}$ & $N_\mathrm{HII}$ & $N_\mathrm{HII, ne}$ & $N_\mathrm{HII,assoc}$  \\
    & \# & \# & \# & \% & \# & \# & \# & \# & \# & \# \\
    & [1] & [2] & [3] & [4] & [5] & [6] & [7] & [8] & [9] & [10] \\
 \hline
    IC~5332 & 786 & 756 & 37 & 4.9 & 36 & 1 & 0 & 36 & 12 & 29 \\
    NGC~1087 & 1004 & 995 & 276 & 27.7 & 186 & 63 & 27 & 274 & 127 & 170 \\
    NGC~1300 & 1448 & 1437 & 229 & 15.9 & 155 & 59 & 15 & 216 & 84 & 109 \\
    NGC~1365 & 1409 & 888 & 193 & 21.7 & 105 & 38 & 50 & 141 & 74 & 86 \\
    NGC~1385 & 1023 & 1023 & 308 & 30.1 & 174 & 69 & 65 & 305 & 193 & 190 \\
    NGC~1433 & 1717 & 1018 & 142 & 13.9 & 99 & 27 & 16 & 114 & 34 & 67 \\
    NGC~1512 & 623 & 489 & 87 & 17.8 & 63 & 11 & 13 & 81 & 28 & 55 \\
    NGC~1566 & 2330 & 2049 & 470 & 22.9 & 288 & 110 & 72 & 439 & 242 & 338 \\
    NGC~1672 & 1559 & 1535 & 414 & 27.0 & 249 & 89 & 76 & 385 & 207 & 256 \\
    NGC~2835 & 1068 & 1007 & 188 & 18.7 & 135 & 46 & 7 & 176 & 120 & 128 \\
    NGC~3351 & 1247 & 1046 & 85 & 8.1 & 64 & 13 & 8 & 76 & 31 & 62 \\
    NGC~3627 & 1598 & 1122 & 280 & 25.0 & 187 & 54 & 39 & 262 & 157 & 192 \\
    NGC~4254 & 2940 & 2464 & 762 & 30.9 & 482 & 212 & 68 & 742 & 396 & 532 \\
    NGC~4303 & 3029 & 2650 & 702 & 26.5 & 438 & 180 & 84 & 674 & 385 & 465 \\
    NGC~4321 & 1806 & 1265 & 327 & 25.8 & 214 & 68 & 45 & 308 & 145 & 230 \\
    NGC~4535 & 1891 & 1523 & 204 & 13.4 & 162 & 35 & 7 & 193 & 77 & 114 \\
    NGC~5068 & 1781 & 1699 & 125 & 7.4 & 104 & 17 & 4 & 123 & 98 & 103 \\
    NGC~628 & 2773 & 2216 & 217 & 9.8 & 166 & 44 & 7 & 210 & 83 & 143 \\
    NGC~7496 & 758 & 728 & 131 & 18.0 & 79 & 33 & 19 & 127 & 51 & 80 \\
    Total & 30790 & 25910 & 5177 & 20.0 & 3386 & 1169 & 622 & 4882 & 2544 & 3349 \\
\hline\hline
\end{tabular}

\tablefoot{
[1] The total number of nebulae within the MUSE catalogue for each galaxy (excluding flagged regions; see \citealp{Groves2023}). 
[2] The total number of MUSE nebulae catalogue within the HST coverage. 
[3] The total number of nebulae within the HST catalogue (excluding flagged regions; see \S\ref{subsec_sourceident}). 
[4] Percentage of regions from the MUSE nebulae catalogue in the HST catalogue.
[5, 6, 7] The total number of nebulae with a complexity scores ($\mathcal{C}$) of $\mathcal{C}\leq1$ (\Csimple), $2\leq\mathcal{C}\leq5$ (\Cintermediate), $\mathcal{C}>5$ (\Ccomplex) within the HST catalogue (see \S\,\ref{subsec_complexity}). 
[8] The total number of nebulae classified as \hii\ regions within the HST catalogue. 
[9] The total number of nebulae classified as \hii\ regions with reliable \ne\ estimates within the HST catalogue (see \S\,\ref{subsec_size_strom}). 
[10] The total number of nebulae classified as \hii\ regions within the HST catalogue, and that have stellar properties from the NUV-selected multi-scale stellar
associations at a scale of 32\,pc (see \S\,\ref{subsec_stellarprops}).
}

\end{table*}

%% file: figures/fig_completeness.tex
\begin{figure}
    \centering
	\includegraphics[width=\columnwidth]{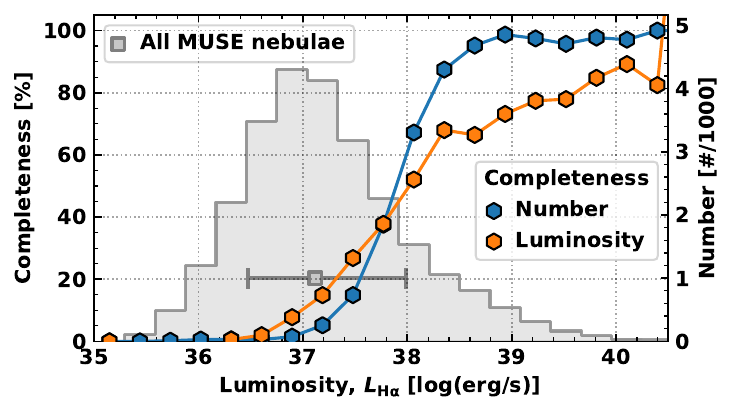}  \\ \vspace{-2mm}
    \caption{\textbf{Catalogue completeness.} Shown is the number (also see Tab.\,\ref{tab_map_compprops}) and luminosity completeness for the HST catalogue with respect to the MUSE catalogue as a function of luminosity, log($L_\mathrm{H\alpha}$). Also shown as a grey histogram (second y-axis) is the total number of regions within the MUSE catalogue within each luminosity bin, where the grey point with error bars shows the median and standard deviation of the distribution.}
    \label{fig_completeness}
\end{figure}

%% file: figures/fig_histrad.tex

\begin{figure}
    \centering
	\includegraphics[width=\columnwidth]{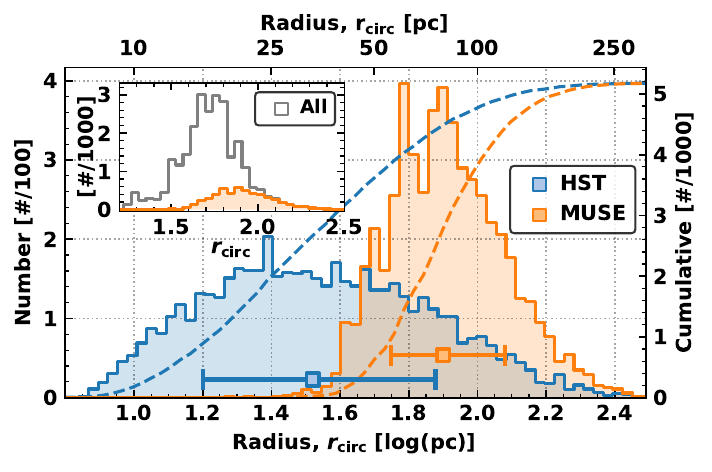}  \\ \vspace{-2mm}
    \caption{\textbf{Distribution of source sizes for all galaxies in the PHANGS-MUSE and PHANGS-MUSE/HST-\ha\ Nebula Catalogue.} The histogram distributions of source radii for galaxies identified in both the HST and MUSE observations are shown in blue and orange, respectively. The points with error bars denote the median and standard deviation of the distributions. It is important to note that only sources detected by both HST and MUSE are displayed in the main panel, excluding the full MUSE sample identified by \citet{Groves2023}. \textcolor{\fontcolor}{The distributions extend down to radii of approximately $5$\,pc, corresponding to the half-width at half-maximum (HWHM) of the assumed Gaussian point spread function (PSF) used to homogenise the observations (see Sect.~\ref{subsec_DataHomogenization}).} The inset panel presents the distribution of source radii for the full MUSE catalogue (restricted to the HST field of view), alongside those detected in the HST catalogue.}
    \label{fig_histrad}
\end{figure}

%% file: figures/fig_histrad_ratio.tex
\begin{figure}
    \centering
        \includegraphics[width=\columnwidth]{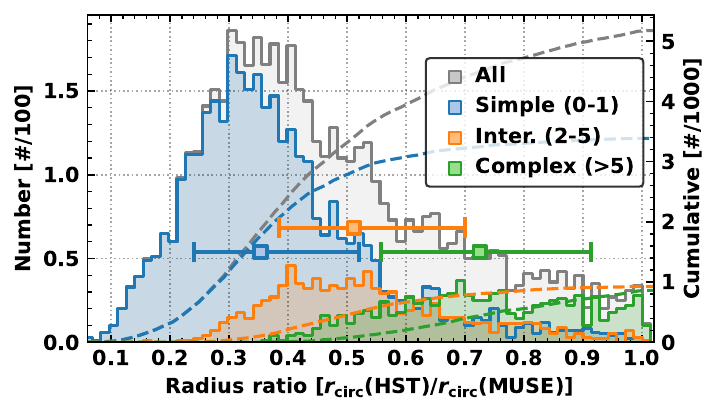} \vspace{-2mm}
    \caption{\textbf{Distribution of source sizes ratios for all galaxies in the Nebula Catalogues relative to MUSE.} The ratio of the size of each region identified in both the HST or MUSE observations for all nebulae in the sample is shown in grey. The ratio of the sizes split between complexity scores of 0-1, 2-5, and >5, denoting simple and intermediate and complex regions, respectively are shown as coloured histograms (see \S\,\ref{subsec_structure}).}
    \label{fig_histrad_ratio}
\end{figure}

%% file: figures/fig_histradratioenv.tex
\begin{figure*}
    \centering
        \includegraphics[width=\textwidth]{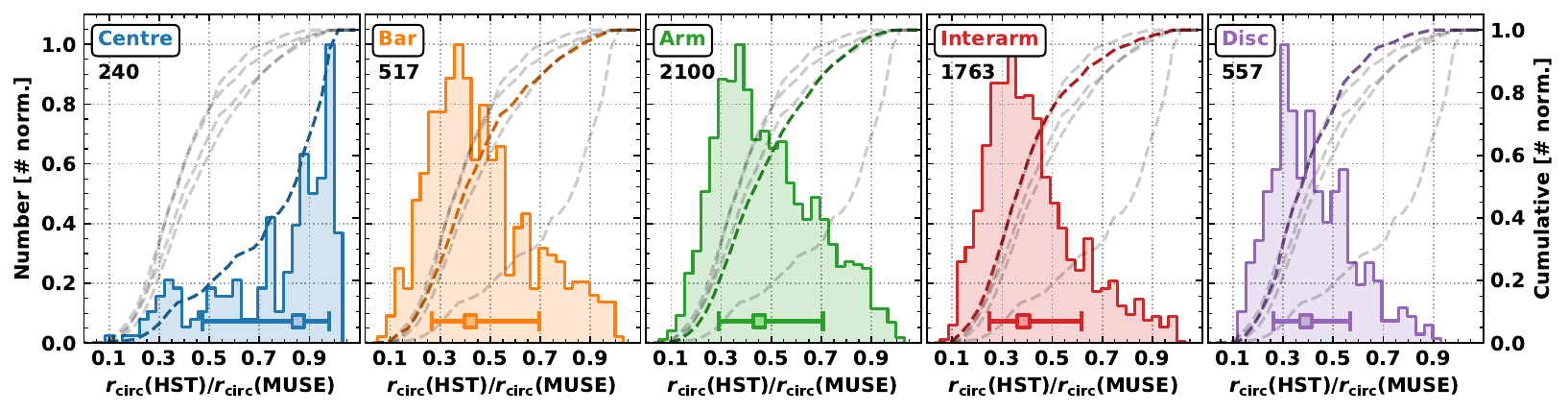}
    \caption{\textbf{Distribution of source size ratios (HST/MUSE) in the nebula catalogue within each environment \citep{Querejeta2021}.} Shown is the ratio of the size of each region identified in both the HST or MUSE observations (see Fig.\,\ref{fig_histrad} for full distribution, and see Fig.\,\ref{fig_histradenv_appendix} for distribution of $r_\mathrm{circ}$ for each environment). We show the histogram and cumulative distributions as solid filled and dashed lines, respectively (see upper left number of regions in each histogram). For comparison, all distributions are normalised to unity, and overlaid on each panel as light dashed grey lines are the cumulative distributions from the other panels.}
    \label{fig_histradenv}
\end{figure*}

%% file: figures/fig_hist_literature.tex
\begin{figure}
    \centering
    \includegraphics[width=\columnwidth]{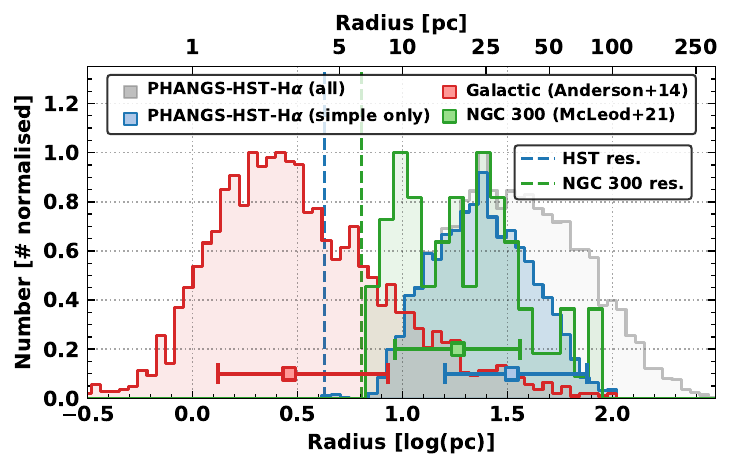}
    \caption{\textbf{Comparison of radius distributions across samples.} Normalised histogram of nebular ($r_\mathrm{circ}$) radii (log scale) for the PHANGS-HST-H$\alpha$ catalogue (\Csimple\ regions only are shown in filled blue, whilst the histogram for all regions is shown in grey), compared with Galactic \hii\ region data from \citet{Anderson2014} and observations of NGC~300 from \citet{McLeod2021}. Vertical dashed lines indicate the approximate resolution limits of the HST (blue) and NGC~300 (green) datasets.}
    \label{fig_hist_lit}
\end{figure}

%% file: figures/fig_histradmoms.tex
\begin{figure}
    \centering
	\includegraphics[width=\columnwidth]{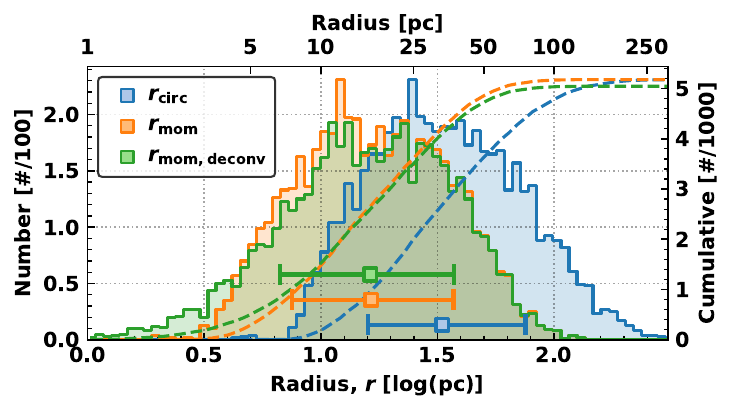} \\ \vspace{-3mm}
    \caption{\textbf{Comparison of the sizes obtained by the circular and the moment methods (see \S\,\ref{subsec_size}).} We show the histogram and the cumulative distribution for both the circular ($r_\mathrm{circ}$; blue), the moment definition of the radius ($r_\mathrm{mom}$; orange) and the deconvolved moment ($r_\mathrm{mom,deconv}$; green).}
    \label{fig_histradmoms}
\end{figure}


%% file: figures/fig_lumrad.tex

\begin{figure*}
    \centering
        \includegraphics[width=\textwidth]{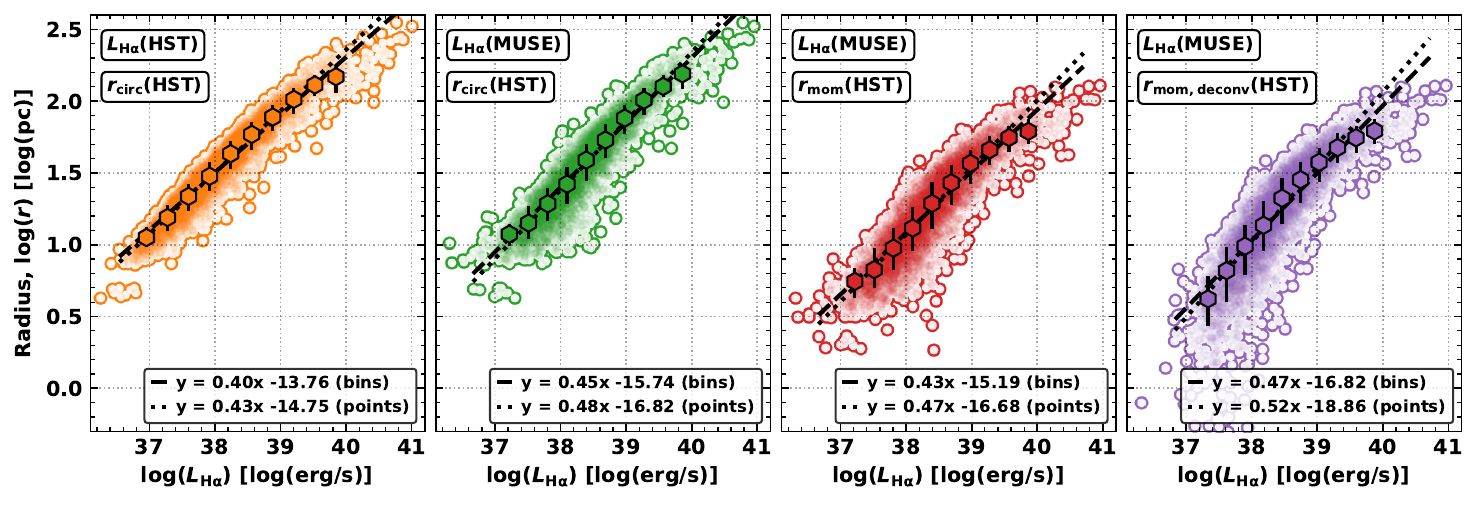}
    \caption{\textbf{Luminosity-size (log-log) distribution for all galaxies.} The distributions of circular radius ($r_\mathrm{circ}$) as a function of luminosity, $L_\mathrm{H\alpha}$(HST), measured in the HST images (first panel). The circular ($r_\mathrm{circ}$; second panel), moment ($r_\mathrm{mom}$; third panel), deconvolved ($r_\mathrm{mom, deconv}$; forth panel) radius from the HST catalogue as a function the luminosity, $L_\mathrm{H\alpha}$(MUSE), from the MUSE catalogue. Overlaid are equally spaced binned points (median values of bins shown), with error bars indicating the standard deviation of the points within each bin. Also overlaid are the results of the fitting for the bins and points between the 1 and 99 percentile ranges in both $L_\mathrm{H\alpha}$ and $r$ (see legend and text for more details).}
    \label{fig_lumrad_fitting}
\end{figure*}

%% file: figures/fig_rad_strom.tex
\begin{figure}
    \centering
    \includegraphics[width=\columnwidth]{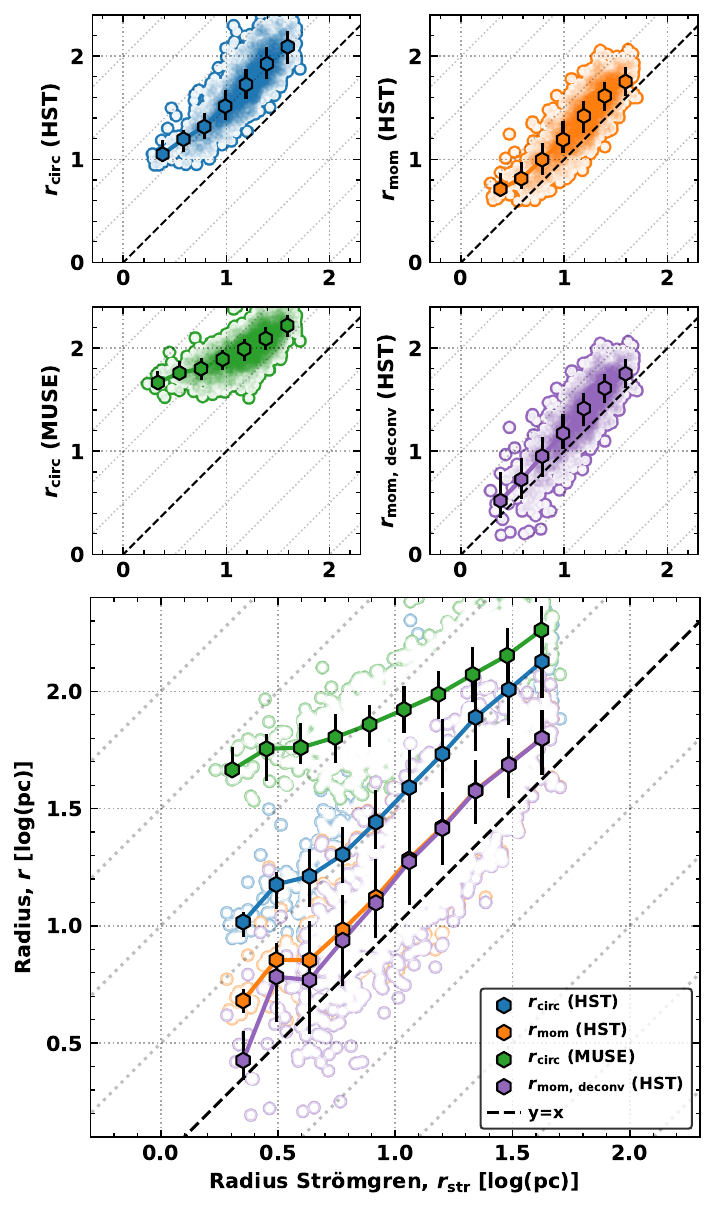}
    \caption{\textbf{Comparison of observed nebula sizes with theoretical Strömgren radii (\S\,\ref{subsec_size_strom}).}
    The top four panels (2\,$\times$\,2) show, individually, $r$ versus the Strömgren radius $r_\mathrm{str}$ for
    $r_\mathrm{circ}$ (HST), $r_\mathrm{mom}$ (HST), $r_\mathrm{circ}$ (MUSE), and $r_\mathrm{mom,\,deconv}$ (HST), respectively.
    The bottom panel overlays all four measurements. In every panel, points are binned at equal intervals in $r_\mathrm{str}$;
    symbols mark bin medians and error bars indicate the standard deviation of values within each bin. The $y=x$ relation is shown
    as a dashed black line (with faint dotted offset guides in the overlay).}
    \label{fig_rad_strom}
\end{figure}

%% file: figures/fig_fillingfraction.tex
\begin{figure}
    \centering
	\includegraphics[width=\columnwidth]{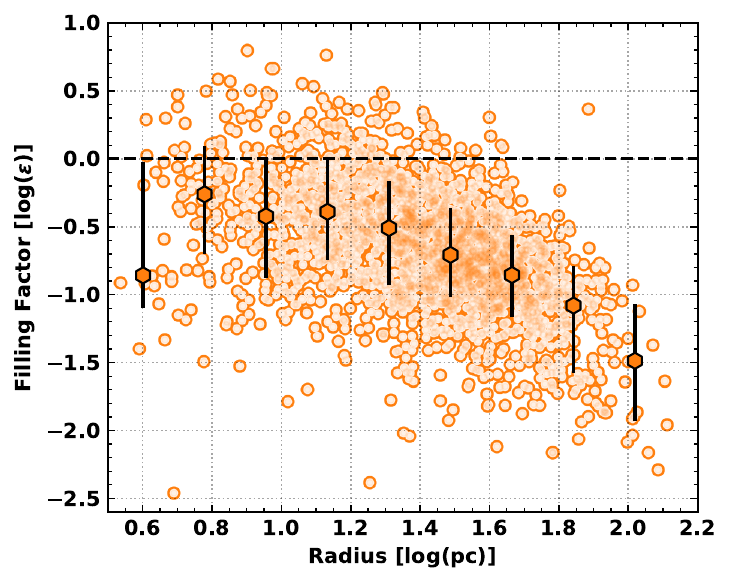}
    \caption{\textbf{Filling factors as a function of nebular size.} 
    The filling factor is computed from the Strömgren balance (Eq.\,\ref{equ_fillingfactor}). 
    Black hexagons with error bars show the median and $\pm1\sigma$ scatter in size bins. 
    The dashed line marks $\epsilon=1$, expected for a uniform-density Strömgren sphere. 
    The systematically low values highlight the clumpy nature of the ionised gas, with [S\textsc{ii}] tracing dense structures embedded within more extended nebulae.}
    \label{fig_rad_filling}
\end{figure}

%% file: figures/fig_histassociations_agemass.tex
\begin{figure}
    \centering
        \includegraphics[width=\columnwidth]{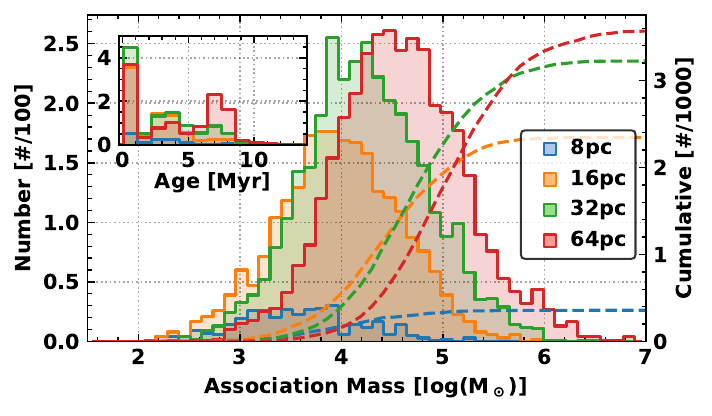}
    \caption{\textbf{Distribution of NUV-selected multi-scale stellar associations connected with the HST Nebula Catalogue (see \S\,\ref{subsec_stellarprops}).} We show the histogram and the cumulative distribution for the stellar association masses (\textit{main panel}) and ages (\textit{inset panel}) as solid filled and dashed lines, respectively. Distributions are shown for stellar associations identified at scales of 8, 16, 32, and 64\,pc (see \citealp{Larson2023}), including \HII\ regions with both single and multiple associations. The 8\,pc scale is not available for galaxies with distances larger than 18\,Mpc (see Tab.\,\ref{tab_galprops}). \textcolor{\fontcolor}{Here, the ``scale'' refers to the Gaussian-smoothing lengths used in the multi-scale watershed segmentation of \citet{Larson2023}, and should not be confused with the physical radii of the \hii\ regions measured in this work.}}
    \label{fig_hist_assoc}
\end{figure}

%% file: figures/fig_rad_mass.tex

\begin{figure}
    \centering
	\includegraphics[width=\columnwidth]{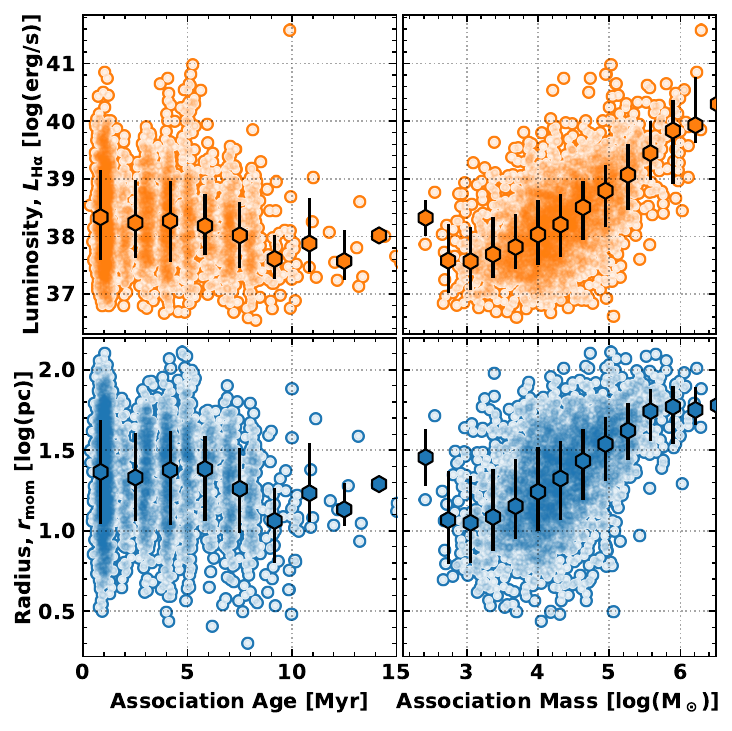} \\ \vspace{-3mm}
    \caption{\textbf{Size and luminosity of the \HII\ regions as a function of the stellar association age (left panels) and mass (right panels).} The top row shows the attenuation-corrected H$\alpha$ luminosity, while the bottom row shows the circularized radius. We show only the NUV-selected multi-scale stellar associations at a scale of 32\,pc, which is comparable to \citet{Scheuermann2023}. Overlaid are equally spaced binned points (median values per bin), with error bars indicating the 16th to 84th percentile range of the data within each bin.}
    \label{fig_rad_age_mass_assoc}
\end{figure}

%% file: figures/fig_bpt.tex
\begin{figure*}
    \centering
	\includegraphics[width=\textwidth]{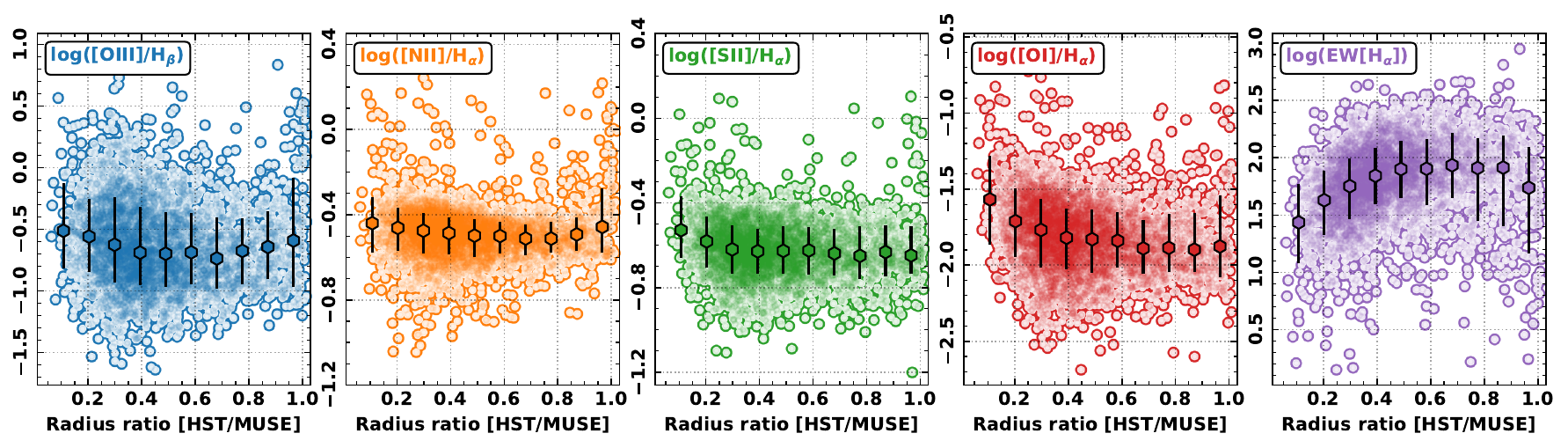}
    \caption{\textbf{Line ratio diagnostic diagrams and radius ratio $r_{\mathrm{HST}}/r_{\mathrm{MUSE}}$ analysis.} Scatter plots of circular ($r_\mathrm{circ}$) radius ratios (HST/MUSE) versus various line diagnostics — [OIII]/H$\beta$, [NII]/H$\alpha$, [SII]/H$\alpha$, and [OI]/H$\alpha$ — along with the equivalent width of H$\alpha$ (in units of log($\AA$)). Overlaid are equally spaced binned points (median values of bins shown), with error bars indicating the standard deviation of the points within each bin.}
    \label{fig_bpt}
\end{figure*}

%% file: tables/table_catalogue.tex
\begin{sidewaystable*}
\centering
\caption{Example entries from the PHANGS-HST H$\alpha$ Nebulae Catalogue.}
\begin{tabular}{cccccccccc}
\hline \hline
\texttt{galaxy\_name} & \texttt{region\_ID} & \texttt{ra} & \texttt{dec} & \texttt{radius\_circ\_pc} & \texttt{halpha\_flux} & \texttt{ne\_sii\_muse} & \texttt{age\_assoc\_nuv\_08pc} & \texttt{mass\_assoc\_nuv\_08pc} & \dots \\
 &  & $\mathrm{{}^{\circ}}$ & $\mathrm{{}^{\circ}}$ & $\mathrm{pc}$ & $\mathrm{erg\,s^{-1}\,cm^{-2}}$ & $\mathrm{cm^{-3}}$ & $\mathrm{Myr}$ & $\mathrm{M_{\odot}}$ & \dots \\
 \hline
IC5332 & 7.0 & 353.606 & -36.115 & 11.88 & 8.07e-16 & 57.52 &  &  & \dots \\
IC5332 & 17.0 & 353.632 & -36.087 & 18.73 & 2.05e-15 & 37.12 &  &  & \dots \\
IC5332 & 26.0 & 353.601 & -36.107 & 27.57 & 8.01e-15 & 85.32 &  & & \dots \\
IC5332 & 28.0 & 353.608 & -36.112 & 16.04 & 1.34e-15 &  &  & & \dots \\
IC5332 & 29.0 & 353.623 & -36.089 & 16.80 & 1.22e-15 &  & 3.00 & 685.84 & \dots\\
IC5332 & 31.0 & 353.595 & -36.084 & 18.96 & 2.65e-15 & 48.27 & 2.00 & 2947.84 & \dots\\
IC5332 & 33.0 & 353.616 & -36.103 & 9.05 & 3.94e-16 &  &  &  & \dots\\
IC5332 & 55.0 & 353.629 & -36.096 & 30.64 & 3.52e-15 &  &  &  & \dots\\
IC5332 & 57.0 & 353.615 & -36.096 & 16.04 & 1.16e-15 & 73.19 &  &  & \dots\\
IC5332 & 60.0 & 353.614 & -36.103 & 19.38 & 1.69e-15 &  & 3.00 & 811.32 & \dots\\
IC5332 & 82.0 & 353.615 & -36.106 & 27.20 & 2.84e-15 &  & 1.00 & 4344.19 & \dots\\
IC5332 & 84.0 & 353.603 & -36.114 & 31.34 & 5.68e-15 & 34.76 & 4.00 & 187.42 & \dots\\
IC5332 & 91.0 & 353.605 & -36.104 & 10.74 & 7.08e-14 &  &  &  & \dots\\
IC5332 & 92.0 & 353.598 & -36.094 & 27.46 & 3.17e-15 &  & 1.00 & 1667.13 & \dots\\
IC5332 & 108.0 & 353.616 & -36.097 & 19.26 & 1.4e-15 &  &  & & \dots \\
IC5332 & 125.0 & 353.611 & -36.082 & 19.38 & 1.72e-15 &  &  & & \dots \\
IC5332 & 127.0 & 353.609 & -36.090 & 14.35 & 9.14e-16 &  &  & & \dots \\
IC5332 & 129.0 & 353.616 & -36.099 & 9.42 & 6.04e-16 & 60.09 &  & & \dots \\
IC5332 & 141.0 & 353.612 & -36.111 & 47.20 & 1.71e-14 & 40.51 & 3.00 & 160.10 & \dots \\
IC5332 & 153.0 & 353.627 & -36.104 & 13.21 & 7.63e-16 &  &  & & \dots \\
\dots & \dots & \dots & \dots & \dots & \dots & \dots & \dots & \dots & \dots \\
\hline
\end{tabular}
\label{tab_ned_cat}
\tablefoot{We show a truncated subset of the full catalogue, including only the first rows for IC\,5332 and a small selection of columns. Displayed parameters include the galaxy name, region identifier, J2000 right ascension and declination of the region centroid, circularized radius (pc), H$\alpha$ flux measured from HST imaging, electron density from the [S\,II] $\lambda6716/\lambda6731$ ratio measured with PHANGS-MUSE, and age and stellar mass of the nearest NUV-selected stellar association at the 8\,pc scale \citep{Larson2023}. The full catalogue contains \Nhstnoflags\ entries across 19 galaxies, with 245 columns providing coordinates, sizes, fluxes, luminosities, morphologies, derived physical parameters, and cross-matched stellar association properties. The complete version of this table is available online at CDS.}
\end{sidewaystable*}

%% file: tables/table_catalogue_descriptions.tex
\begin{table*}
\centering
\caption{Description of selected columns from the PHANGS-MUSE/HST H$\alpha$
Nebulae Catalogue (part I).}
\begin{tabular}{cc}
\hline \hline
Column & Description \\
\hline

\texttt{galaxy\_name} & Galaxy name \\
\texttt{region\_ID} & Region identifier \\
\texttt{flag\_edge\_hst} & Flag=1 if region lies on HST image edge \\
\texttt{flag\_touch\_hst} & Flag=1 if region touches segmentation boundary \\
\texttt{flag\_manual\_hst} & Manual classification flags (visual checks) \\
\texttt{flag\_edge\_muse} & Flag=1 if region lies on MUSE cube edge \\
\texttt{flag\_star\_muse} & Flag=1 if affected by bright star in MUSE data \\
\texttt{ra} & Right ascension (J2000) of region centroid \\
\texttt{dec} & Declination (J2000) of region centroid \\
\texttt{ra\_peak} & Right ascension (J2000) of \ha\ peak pixel \\
\texttt{dec\_peak} & Declination (J2000) of \ha\ peak pixel \\
\texttt{ra\_muse} & Right ascension (J2000) of MUSE centroid \\
\texttt{dec\_muse} & Declination (J2000) of MUSE centroid \\
\texttt{r\_R25\_deproj} & Deprojected galactocentric radius, in R25 units \\
\texttt{r\_reff\_deproj} & Deprojected galactocentric radius, in Re units \\
\texttt{phi\_deproj} & Deprojected azimuthal angle within host galaxy \\
\texttt{environment} & Environment flag following \citet{Querejeta2021} \\
\texttt{area\_circ\_arcsec2} & Circularized area (arcsec$^{2}$) \\
\texttt{radius\_circ\_arcsec} & Circularized radius (arcsec) \\
\texttt{radius\_circ\_pc} & Circularized radius (pc) \\
\texttt{area\_mom\_arcsec2} & Second-moment area of region (arcsec$^{2}$) \\
\texttt{radius\_major\_mom\_arcsec} & Second-moment major axis (arcsec) \\
\texttt{radius\_minor\_mom\_arcsec} & Second-moment minor axis (arcsec) \\
\texttt{radius\_pos\_angle\_deg} & Position angle (moments-based) \\
\texttt{radius\_mom\_arcsec} & Moment-based radius (arcsec) \\
\texttt{radius\_mom\_pc} & Moment-based radius (pc) \\
\texttt{radius\_mom\_deconv\_pc} & Deconvolved moment-based radius (pc) \\
\texttt{radius\_stromgren\_pc} & Stromgren radius (pc) \\
\texttt{radius\_circ\_pc\_muse} & Circularized radius (pc; PHANGS-MUSE) \\
\texttt{complexity\_score} & Morphological complexity score \\
\texttt{halpha\_flux} & HST \ha\ integrated flux (raw, uncorrected) \\
\texttt{halpha\_flux\_err} & Error on HST \ha\ flux \\
\texttt{halpha\_flux\_max} & Maximum pixel flux in HST \ha\ map \\
\texttt{halpha\_flux\_min} & Minimum pixel flux in HST \ha\ map \\
\texttt{halpha\_flux\_mean} & Mean pixel flux in HST \ha\ map \\
\texttt{halpha\_flux\_corr} & Extinction-corrected HST \ha\ flux \\
\texttt{halpha\_flux\_corr\_err} & Error on extinction-corrected HST \ha\ flux \\
\texttt{<line>\_flux\_muse} & Measured emission-line flux (PHANGS-MUSE) \\
\texttt{<line>\_flux\_err\_muse} & Statistical uncertainty on flux (PHANGS-MUSE) \\
\texttt{<line>\_flux\_corr\_muse} & Extinction-corrected flux (PHANGS-MUSE) \\
\texttt{<line>\_flux\_corr\_err\_muse} & Uncertainty on corrected flux (PHANGS-MUSE) \\
\texttt{<line>\_vel\_muse} & Velocity of <line> (PHANGS-MUSE) \\
\texttt{<line>\_vel\_err\_muse} & Uncertainty on velocity of <line> (PHANGS-MUSE) \\
\texttt{<line>\_sigma\_muse} & Velocity dispersion of <line> (PHANGS-MUSE) \\
\texttt{<line>\_sigma\_err\_muse} & Uncertainty on velocity dispersion of <line> (PHANGS-MUSE) \\
\texttt{<line>\_ew\_muse} & Equivalent width of <line> (Å; PHANGS-MUSE) \\
\texttt{<line>\_ew\_err\_muse} & Uncertainty on equivalent width of <line> (Å; PHANGS-MUSE) \\
\texttt{<line>\_ew\_<method>\_muse} & Equivalent width of <line> using <method> (Å; PHANGS-MUSE) \\
\texttt{<line>\_ew\_<method>\_err\_muse} & Uncertainty on equivalent width of <line> using <method> (Å; PHANGS-MUSE) \\

\hline
\end{tabular}
\tablefoot{This table lists structural, positional, and
spectroscopic parameters, including fluxes, velocities, dispersions, and
equivalent widths. The parameters correspond to those shown in the example
catalogue entries in Tab.\,\ref{tab_ned_cat}. Here \texttt{<line>} denotes an emission line (e.g. \texttt{halpha}, \texttt{hbeta}, \texttt{oiii5007}, \texttt{nii6584}, \texttt{sii6716}, \texttt{sii6731}, \texttt{siii9069}, \texttt{oi6300}, \texttt{nii5755}, \texttt{hei5876}, \texttt{siii6312}, \texttt{oi6363}, \texttt{oii7319}, \texttt{oii7330}). All fluxes are in erg\,s$^{-1}$\,cm$^{-2}$; velocities and dispersions are in km\,s$^{-1}$; equivalent widths are in Å. The suffix \texttt{\_err} indicates 1$\sigma$ statistical uncertainties; \texttt{\_corr} indicates extinction-corrected values. The suffixes \texttt{\_vel} and \texttt{\_sigma} give line-of-sight velocity and velocity dispersion, respectively. EW variants use \texttt{\_raw} (direct from the spectrum), \texttt{\_fit} (from a fitted continuum/emission model), and \texttt{\_bgcorr} (background-corrected). HST-specific columns \texttt{halpha\_flux\_max}, \texttt{\_min}, and \texttt{\_mean} are pixel statistics from the HST H$\alpha$ map; \texttt{halpha\_flux\_corr} and \texttt{\_corr\_err} are extinction-corrected flux and its uncertainty.}
\label{tab_ned_cat_desc1}
\end{table*}

\begin{table*}
\centering
\caption{Description of selected columns from the PHANGS-MUSE/HST H$\alpha$
Nebulae Catalogue (part II, continued).}
\begin{tabular}{cc}
\hline \hline
Column & Description \\
\hline
\texttt{bpt\_nii\_muse} & BPT class using [N II]/\ha\ (PHANGS-MUSE) \\
\texttt{bpt\_sii\_muse} & BPT class using [S II]/\ha\ (PHANGS-MUSE) \\
\texttt{bpt\_oi\_muse} & BPT class using [O I]/\ha\ (PHANGS-MUSE) \\
\texttt{halpha\_lum} & \ha\ luminosity from HST (erg/s) \\
\texttt{halpha\_lum\_err} & Uncertainty on HST \ha\ luminosity (erg/s) \\
\texttt{halpha\_lum\_muse} & \ha\ luminosity from PHANGS-MUSE (erg/s; PHANGS-MUSE) \\
\texttt{hii\_class\_v2\_muse} & PHANGS-MUSE H II classification (v2; PHANGS-MUSE) \\
\texttt{hii\_class\_v3\_muse} & PHANGS-MUSE H II classification (v3; PHANGS-MUSE) \\
\texttt{te\_nii\_muse} & Electron temperature from [N II] (K; PHANGS-MUSE) \\
\texttt{te\_nii\_err\_muse} & Uncertainty on Te([N II]) (K; PHANGS-MUSE) \\
\texttt{te\_siii\_muse} & Electron temperature from [S III] (K; PHANGS-MUSE) \\
\texttt{te\_siii\_err\_muse} & Uncertainty on Te([S III]) (K; PHANGS-MUSE) \\
\texttt{ne\_sii\_muse} & Electron density from [S II] 6716/6731 (cm$^{-3}$; PHANGS-MUSE) \\
\texttt{ne\_sii\_err\_muse} & Uncertainty on ne([S II]) (cm$^{-3}$; PHANGS-MUSE) \\
\texttt{q} & Ionizing photon rate $Q$ (s$^{-1}$) \\
\texttt{q\_err} & Uncertainty on $Q$ (s$^{-1}$) \\
\texttt{logq\_d91\_muse} & log10(ionization parameter) (PHANGS-MUSE) \\
\texttt{logq\_d91\_err\_muse} & Uncertainty on log10(ionization parameter) (PHANGS-MUSE) \\
\texttt{ebv\_muse} & Color excess E(B-V) from Balmer decrement (mag; PHANGS-MUSE) \\
\texttt{ebv\_err\_muse} & Uncertainty on E(B-V) (mag; PHANGS-MUSE) \\
\texttt{av\_muse} & Visual extinction $A_V$ (mag; PHANGS-MUSE) \\
\texttt{metallicity\_scal\_muse} & Gas-phase metallicity (Scal code; PHANGS-MUSE) \\
\texttt{metallicity\_scal\_err\_muse} & Uncertainty on metallicity (Scal; PHANGS-MUSE) \\
\texttt{metallicity\_scal\_delta\_muse} & Offset from local metallicity trend (dex; PHANGS-MUSE) \\
\texttt{age\_assoc\_<filter>\_<scale>} & Stellar association age (Myr; PHANGS-HST) \\
\texttt{age\_err\_assoc\_<filter>\_<scale>} & Uncertainty on association age (Myr; PHANGS-HST) \\
\texttt{ra\_assoc\_<filter>\_<scale>} & Association right ascension (deg; PHANGS-HST) \\
\texttt{dec\_assoc\_<filter>\_<scale>} & Association declination (deg; PHANGS-HST) \\
\texttt{ebv\_assoc\_<filter>\_<scale>} & Color excess E(B--V) (mag; PHANGS-HST) \\
\texttt{ebv\_err\_assoc\_<filter>\_<scale>} & Uncertainty on E(B--V) (mag; PHANGS-HST) \\
\texttt{mass\_assoc\_<filter>\_<scale>} & Stellar mass (M$_\odot$; PHANGS-HST) \\
\texttt{mass\_err\_assoc\_<filter>\_<scale>} & Uncertainty on stellar mass (M$_\odot$; PHANGS-HST) \\
\texttt{region\_ID\_assoc\_<filter>\_<scale>} & Association identifier (PHANGS-HST) \\
\texttt{flag\_multi\_assoc\_<filter>\_<scale>} & Flag: multiple matched associations (PHANGS-HST) \\
\texttt{flag\_one\_assoc\_<filter>\_<scale>} & Flag: exactly one matched association (PHANGS-HST) \\
\texttt{flag\_none\_assoc\_<filter>\_<scale>} & Flag: no matched association (PHANGS-HST) \\

\hline
\end{tabular}
\tablefoot{This table provides additional
parameters, including luminosities, physical conditions, ionisation properties,
and stellar association matches. Together with
Tab.\,\ref{tab_ned_cat_desc1}, this forms a full overview of the
catalogue description. Columns related to stellar associations follow a repeated naming pattern. The format is \texttt{<quantity>\_assoc\_<filter>\_<scale>}, where \texttt{<filter>} indicates the photometric band used for association identification (\texttt{nuv} = near-ultraviolet; \texttt{v} = optical $V$-band), and \texttt{<scale>} specifies the physical association scale in parsecs (\texttt{08pc}, \texttt{16pc}, \texttt{32pc}, \texttt{64pc}). For example, \texttt{age\_assoc\_nuv\_08pc} denotes the age of the stellar association identified in the NUV at an 8~pc scale, while \texttt{mass\_assoc\_v\_32pc} would give the stellar mass of the $V$-band association at a 32~pc scale.}

\label{tab_ned_cat_desc2}
\end{table*}

%% file: tables/tab_galprops.tex
\begin{table*}
    \centering
    \caption{Properties of the galaxy sample.}
    \label{tab_galprops}
    
    \begin{tabular}{ccccccccccc}
    \hline\hline
   Galaxy & $i$ & PA & Morph. & Dist. & $R_\mathrm{eff}$ & Metal. & $M_\mathrm{HI}$ & $M_\mathrm{H_{2}}$ & $M_\star$ & SFR \\
 & $\mathrm{{}^{\circ}}$ & $\mathrm{{}^{\circ}}$ & & $\mathrm{Mpc}$ & $\mathrm{kpc}$ & 12+log(O/H) & log($\mathrm{M_{\odot}}$) & log($\mathrm{M_{\odot}}$) & log($\mathrm{M_{\odot}}$) & log($\mathrm{M_{\odot}\,yr^{-1}}$) \\
  & $(a)$ & $(a)$ & $(b)$ & $(c)$ & $(d)$ & $(e)$ & $(f)$ & $(g)$ & $(h)$ & $(h)$ \\
 \hline 

IC\,5332 & 26.9 & 74.4 & SABc & 9.0 & 3.6 & 8.37 & 9.3 & -- & 9.7 & -0.4 \\
NGC\,0628* & 8.9 & 20.7 & Sc & 9.8 & 3.9 & 8.48 & 9.7 & 9.4 & 10.3 & 0.2 \\
NGC\,1087 & 42.9 & 359.1 & Sc & 15.9 & 3.2 & 8.40 & 9.1 & 9.2 & 9.9 & 0.1 \\
NGC\,1300 & 31.8 & 278.0 & Sbc & 19.0 & 6.5 & 8.52 & 9.4 & 9.4 & 10.6 & 0.1 \\
NGC\,1365 & 55.4 & 201.1 & Sb & 19.6 & 2.8 & 8.53 & 9.9 & 10.3 & 11.0 & 1.2 \\
NGC\,1385 & 44.0 & 181.3 & Sc & 17.2 & 3.4 & 8.42 & 9.2 & 9.2 & 10.0 & 0.3 \\
NGC\,1433 & 28.6 & 199.7 & SBa & 18.6 & 4.3 & 8.54 & 9.4 & 9.3 & 10.9 & 0.1 \\
NGC\,1512 & 42.5 & 261.9 & Sa & 18.8 & 4.8 & 8.55 & 9.9 & 9.1 & 10.7 & 0.1 \\
NGC\,1566 & 29.5 & 214.7 & SABb & 17.7 & 3.2 & 8.55 & 9.8 & 9.7 & 10.8 & 0.7 \\
NGC\,1672 & 42.6 & 134.3 & Sb & 19.4 & 3.4 & 8.54 & 10.2 & 9.9 & 10.7 & 0.9 \\
NGC\,2835 & 41.3 & 1.0 & Sc & 12.2 & 3.3 & 8.38 & 9.5 & 8.8 & 10.0 & 0.1 \\
NGC\,3351 & 45.1 & 193.2 & Sb & 10.0 & 3.0 & 8.59 & 8.9 & 9.1 & 10.4 & 0.1 \\
NGC\,3627 & 57.3 & 173.1 & Sb & 11.3 & 3.6 & 8.55 & 9.1 & 9.8 & 10.8 & 0.6 \\
NGC\,4254 & 34.4 & 68.1 & Sc & 13.1 & 2.4 & 8.53 & 9.5 & 9.9 & 10.4 & 0.5 \\
NGC\,4303 & 23.5 & 312.4 & Sbc & 17.0 & 3.4 & 8.56 & 9.7 & 9.9 & 10.5 & 0.7 \\
NGC\,4321 & 38.5 & 156.2 & SABb & 15.2 & 5.5 & 8.56 & 9.4 & 9.9 & 10.7 & 0.6 \\
NGC\,4535 & 44.7 & 179.7 & Sc & 15.8 & 6.3 & 8.54 & 9.6 & 9.6 & 10.5 & 0.3 \\
NGC\,5068 & 35.7 & 342.4 & Sc & 5.2 & 2.0 & 8.30 & 8.8 & 8.4 & 9.4 & -0.6 \\
NGC\,7496 & 35.9 & 193.7 & Sb & 18.7 & 3.8 & 8.49 & 9.1 & 9.3 & 10.0 & 0.4 \\

    \hline\hline
    \end{tabular}

    \tablefoot{
    We show in columns from left to right the galaxy name, inclination~($i$), position angle~(PA), morphological type (Morph.), distance (Dist.), effective radius ($R_\mathrm{eff}$), globally averaged metallicity ($12+\log(\mathrm{O/H})$), total mass of atomic gas ($M_\mathrm{HI}$), molecular gas ($M_\mathrm{H_{2}}$) and stars ($M_\star$), and global star formation rate~(SFR).
    *We only make use of the archival "central" pointing observed with the F658N ACS HST filter, due to the lack of cross-over between the MUSE and newer F658N WFC3 HST filter "east" pointing (see \citealp{Chandar2025}).  
    $(a)$ From \cite{Lang2020}, based on PHANGS \mbox{CO(2--1)} kinematics. For IC\,5332, we use values from \cite{Salo2015}.
    $(b)$ Morphological classification taken from HyperLEDA \citep{Makarov2014}.
    $(c)$ Source distances are taken from the compilation of \citet{Anand2021a, Anand2021b}.  
    $(d)$ $R_\mathrm{eff}$ that contains half of the stellar mass of the galaxy \citep{Leroy2021a}. 
    $(e)$ Averaged metallicity within the area mapped by MUSE, computed using the Scal method of \citet{Pilyugin2016}; see \citet{Kreckel2019} for more details. 
    $(f)$ Total atomic gas mass taken from HYPERLEDA \citep{Makarov2014}.
    $(g)$ Molecular gas mass determined from PHANGS \mbox{CO(2--1)} observations (see \citealp{Leroy2021a}). CO was not detected at high enough significance in IC\,5332 to allow a molecular gas mass to be determined.
    $(h)$ Derived by \citet{Leroy2021a}, using \textit{GALEX}~UV and \textit{WISE}~IR photometry, following a similar methodology to \cite{Leroy2019}.
    }
      
\end{table*}

%% file: figures/fig_maps_zoom.tex

\begin{figure*}[!t]
    \centering
        \includegraphics[width=\textwidth]{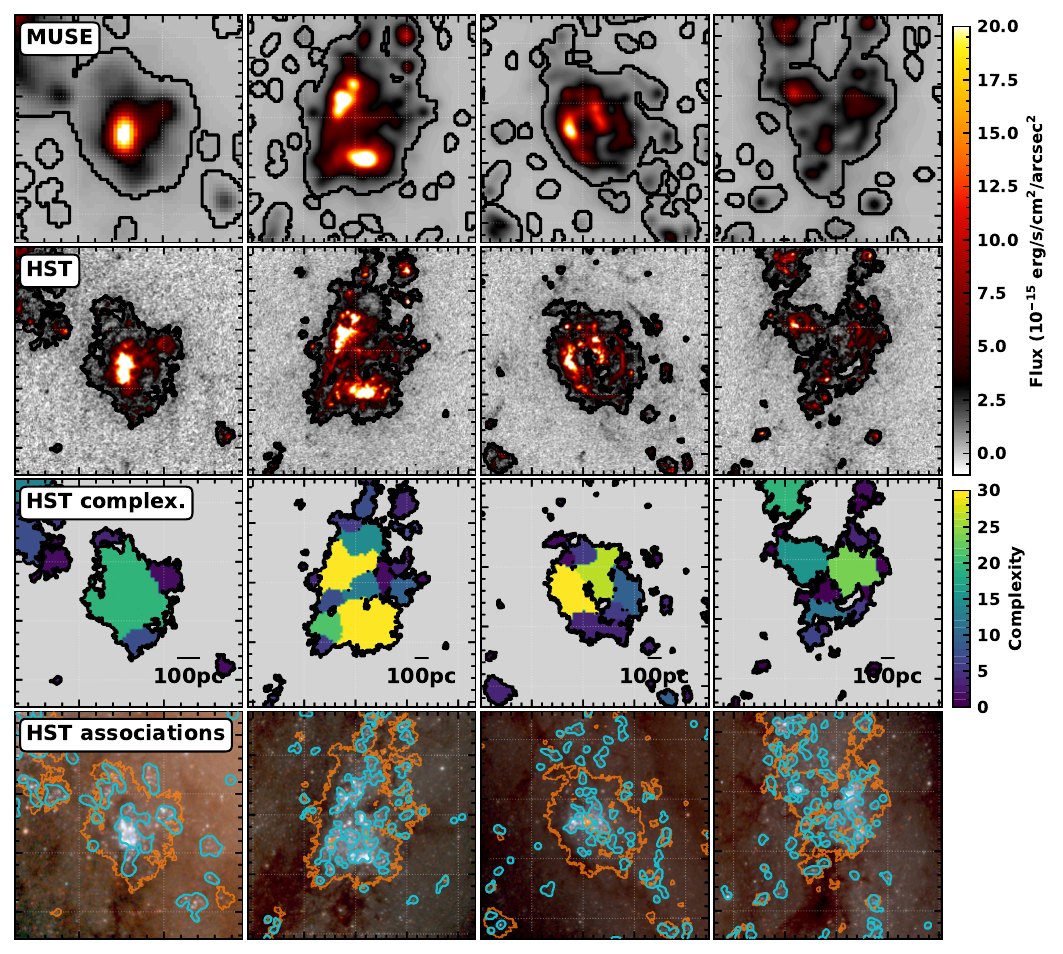}
    \caption{\textbf{Example of several regions within the nebula catalogues towards NGC\,1566.} The MUSE (\textit{first row}) and HST (\textit{second row}) observations, overlaid with contours showing the boundary of each source in the respective observations. (\textit{third row}) We show a map of the complexity score for each region. (\textit{fourth row}) We show the nebula (red contours) and 32\,pc NUV-identified stellar association (\citealp{Larson2023}; blue contour) overlaid on a HST filter red (F814W) green (F555W) blue (F438W+F336W) image (see \citealp{Lee2022}).}
    \label{fig_maps_zoom_complex}
\end{figure*}

%% file: figures/fig_histlum.tex
\begin{figure}
    \centering
        \includegraphics[width=\columnwidth]{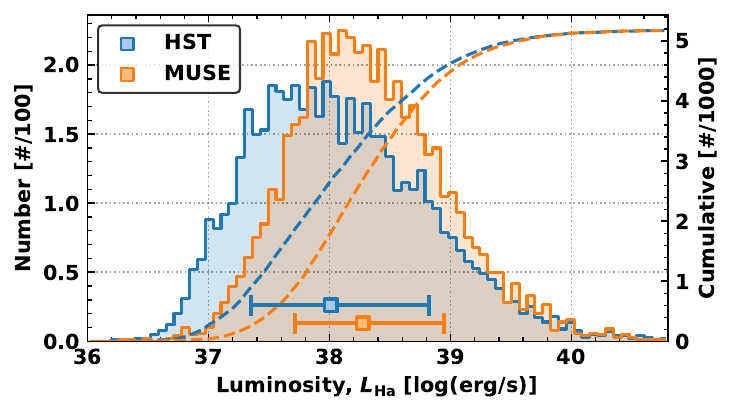} \\ \vspace{-3mm}
        \includegraphics[width=\columnwidth]{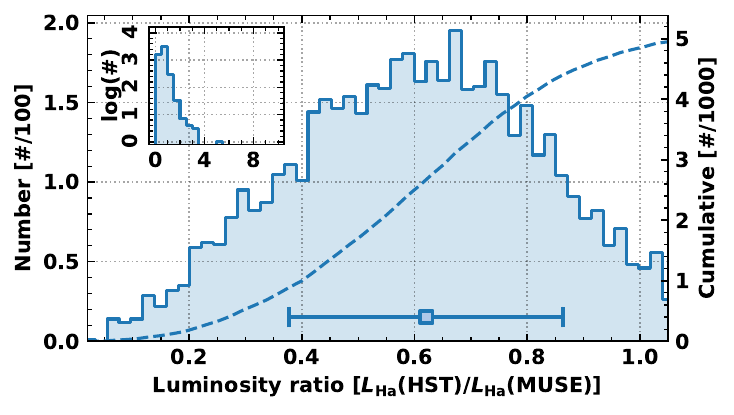}
    \caption{\textbf{Distribution of source luminosities for all regions in the Nebula Catalogues.} In blue and orange we show the distribution for sources identified in both the HST and MUSE observations. Note that we only show the MUSE sample that was also identified by HST (i.e. not the full sample identified by \citealp{Groves2023}). The inset axis shows a larger range in luminosity ratios, and is shown in a log distribution for clarity. In contrast to Fig.\,\ref{fig_histrad}, this shows that despite the regions being much smaller in the HST observations, they still typically retain a significant fraction of the flux.}
    \label{fig_histlum}
\end{figure}

%% file: figures/fig_radlumratios.tex
\begin{figure*}
    \centering
        \includegraphics[width=\textwidth]{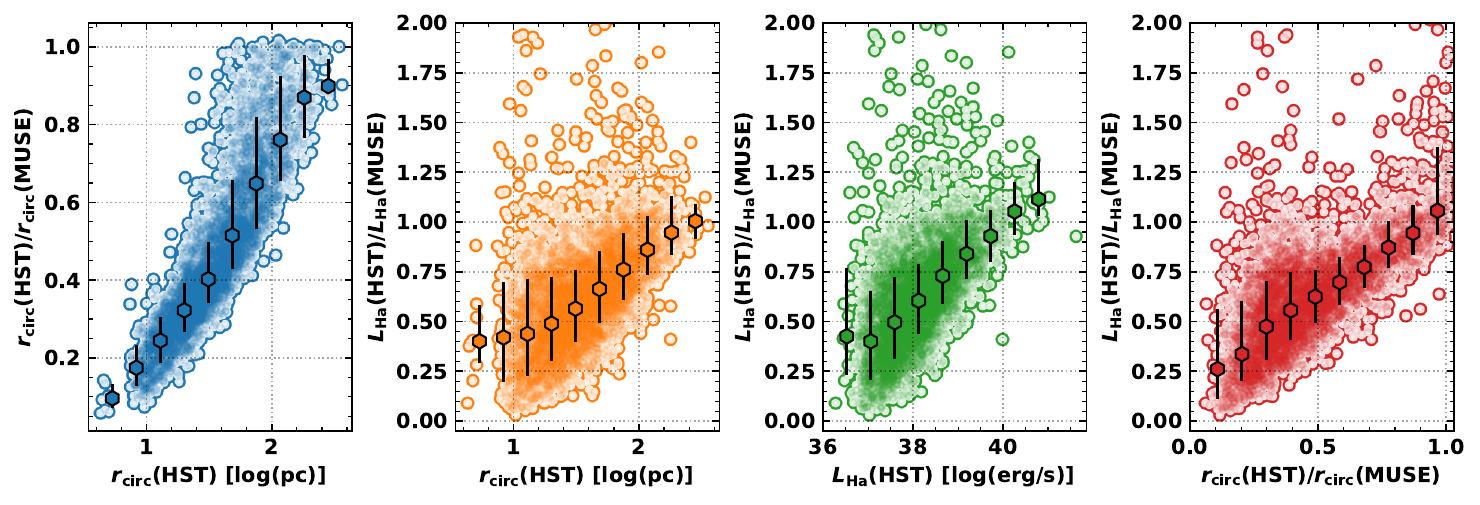}
    \caption{\textbf{Distribution of sizes and luminosities.} (\textit{First panel}) We show the ratio of $r_\mathrm{circ}(\mathrm{HST})/r_\mathrm{circ}(\mathrm{MUSE})$ as a function of the HST sizes ($r_\mathrm{circ}(\mathrm{HST})$). (\textit{Second panel}) We show the luminosity ratio $L_\mathrm{H\alpha}(\mathrm{HST}$)/$L_\mathrm{H\alpha}(\mathrm{MUSE})$ as a function of the HST sizes ($r_\mathrm{circ}(\mathrm{HST})$). (Third panel) We show the luminosity ratio as a function of the HST luminosity, $L_\mathrm{H\alpha}(\mathrm{HST}$). (Fourth panel) We show the luminosity ratio as a function of the radius ratio. Overlaid on all panels are equally spaced binned points (median values of bins shown), with error bars indicating the standard deviation of the points within each bin.}
    \label{fig_radlumratios}
\end{figure*}

%% file: figures/fig_histlumenv.tex
\begin{figure*}
    \centering
        \includegraphics[width=\textwidth]{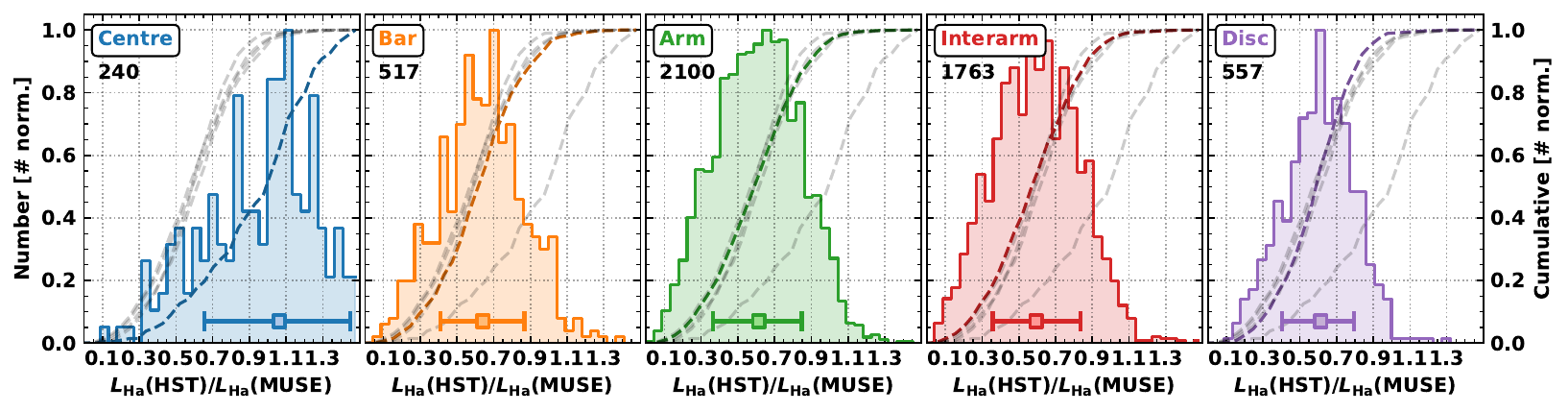}
    \caption{\textbf{Distribution of source luminosities ratios (HST/MUSE) in the nebula catalogue within each environment \citep{Querejeta2021}.} The ratio of the luminosities of each region identified in both the HST or MUSE observations (see lower panel of Fig.\,\ref{fig_histlum} for full distribution). We show the histogram and cumulative distributions as solid filled and dashed lines, respectively. For comparison, all distributions are normalised to unity, and overlaid on each panel as light dashed grey lines are the cumulative distributions from the other panels.}
    \label{fig_histlumenv}
\end{figure*}

%% file: figures/fig_histradenv.tex
\begin{figure*}
    \centering
        \includegraphics[width=\textwidth]{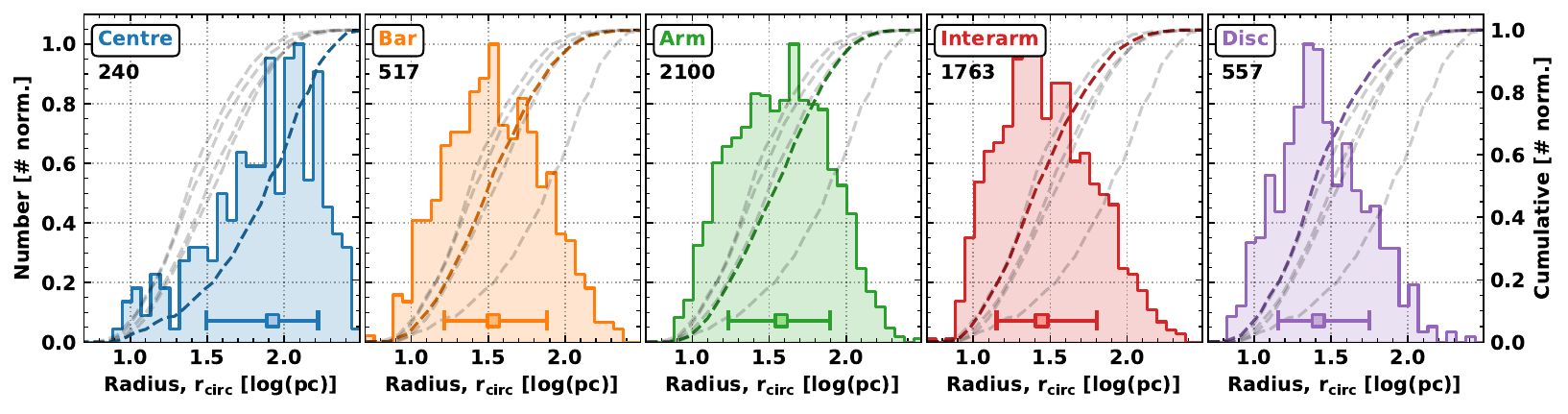}
    \caption{\textbf{Distribution of source sizes in the nebula catalogue within each environment \citep{Querejeta2021}.} We show the histogram and cumulative distributions as solid filled and dashed lines, respectively (see upper left number of regions in each histogram). For comparison, all distributions are normalised to unity, and overlaid on each panel as light dashed grey lines are the cumulative distributions from the other panels.}
    \label{fig_histradenv_appendix}
\end{figure*}

%% file: figures/fig_bpt_lum.tex
\begin{figure*}
    \centering
	\includegraphics[width=\textwidth]{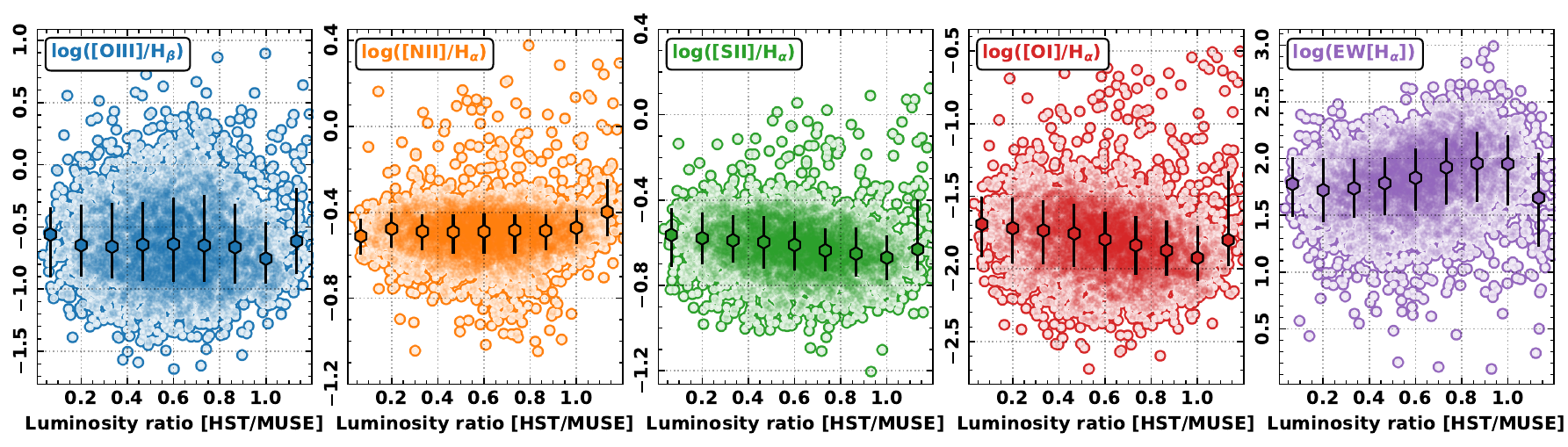}
    \caption{\textbf{Line ratio diagnostic diagrams and luminosity ratio analysis.} Scatter plots of circular ($r_\mathrm{circ}$) radius ratios (HST/MUSE) versus various line diagnostics — [OIII]/H$\beta$, [NII]/H$\alpha$, [SII]/H$\alpha$, and [OI]/H$\alpha$ — along with the equivalent width of H$\alpha$ (in units of log($\AA$)). Overlaid are equally spaced binned points (median values of bins shown), with error bars indicating the standard deviation of the points within each bin.}
    \label{fig_bpt_lum}
\end{figure*}

%% file: tables/tab_radlumstats.tex
\begin{sidewaystable*}
\caption{Distribution of attenuation-corrected H$\alpha$ luminosities and characteristic radii for each galaxy.}
\label{tab_radlum}
\centering
\begin{tabular}{l ccc ccc ccc ccc ccc ccc ccc ccc ccc}
\hline \hline
Galaxy & \multicolumn{3}{c}{L$_\mathrm{H\alpha}$ (HST)} & \multicolumn{3}{c}{L$_\mathrm{H\alpha}$ (MUSE)} & \multicolumn{3}{c}{$r_{\rm circ}$ (HST)} & \multicolumn{3}{c}{$r_{\rm mom}$ (HST)} & \multicolumn{3}{c}{$r_{\rm mom,deconv}$ (HST)} & \multicolumn{3}{c}{$r_{\rm str}$} & \multicolumn{3}{c}{$r_{\rm circ}$ (MUSE)} \\

& \multicolumn{3}{c}{[log(erg s$^{-1}$)]} & \multicolumn{3}{c}{[log(erg s$^{-1}$)]} & \multicolumn{3}{c}{[pc]} & \multicolumn{3}{c}{[pc]} & \multicolumn{3}{c}{[pc]} & \multicolumn{3}{c}{[pc]} & \multicolumn{3}{c}{[pc]} \\
 & 10\% & 50\% & 90\% & 10\% & 50\% & 90\% & 10\% & 50\% & 90\% & 10\% & 50\% & 90\% & 10\% & 50\% & 90\% & 10\% & 50\% & 90\% & 10\% & 50\% & 90\% \\
\hline
\hline

    IC~5332 & 36.9 & 37.4 & 38.0 & 37.1 & 37.7 & 38.2 & 11.1 & 21.0 & 38.2 & 5.3 & 9.7 & 19.4 & 5.1 & 8.9 & 19.2 & 4.1 & 7.2 & 16.5 & 46.0 & 59.1 & 80.7 \\
    NGC~0628 & 37.1 & 37.6 & 38.4 & 37.3 & 37.9 & 38.6 & 11.6 & 21.8 & 47.6 & 5.1 & 10.4 & 23.7 & 3.8 & 9.8 & 23.8 & 6.2 & 13.0 & 22.0 & 40.4 & 60.7 & 95.4 \\
    NGC~1087 & 37.2 & 38.0 & 39.0 & 37.6 & 38.2 & 39.1 & 15.1 & 34.2 & 85.8 & 7.2 & 16.5 & 42.7 & 6.0 & 16.1 & 42.5 & 8.7 & 21.9 & 33.9 & 52.0 & 77.8 & 141.7 \\
    NGC~1300 & 37.2 & 37.8 & 38.8 & 37.5 & 38.1 & 38.8 & 12.5 & 24.4 & 65.6 & 5.8 & 11.7 & 34.0 & 4.3 & 11.2 & 34.0 & 4.7 & 14.4 & 24.5 & 58.8 & 75.6 & 121.7 \\
    NGC~1365 & 37.2 & 38.2 & 40.1 & 37.6 & 38.5 & 40.1 & 12.9 & 40.0 & 153.5 & 5.8 & 19.6 & 71.9 & 4.3 & 19.4 & 72.1 & 8.3 & 22.0 & 35.1 & 77.3 & 96.4 & 183.4 \\
    NGC~1385 & 37.3 & 38.3 & 39.3 & 37.8 & 38.5 & 39.4 & 16.6 & 45.7 & 120.9 & 8.0 & 23.4 & 55.6 & 7.0 & 23.4 & 55.4 & 8.7 & 21.6 & 34.1 & 49.6 & 84.8 & 149.9 \\
    NGC~1433 & 37.0 & 37.7 & 38.7 & 37.3 & 37.9 & 38.7 & 10.5 & 24.5 & 61.6 & 4.9 & 12.0 & 31.9 & 4.3 & 11.8 & 31.8 & 6.2 & 13.8 & 21.7 & 58.5 & 73.5 & 124.8 \\
    NGC~1512 & 36.9 & 37.7 & 39.0 & 37.4 & 38.0 & 39.1 & 11.3 & 25.4 & 102.5 & 5.0 & 11.9 & 51.5 & 3.5 & 11.1 & 51.4 & 5.4 & 16.2 & 24.4 & 79.1 & 85.6 & 177.4 \\
    NGC~1566 & 37.2 & 38.0 & 39.1 & 37.6 & 38.4 & 39.2 & 14.2 & 35.1 & 103.6 & 6.7 & 17.2 & 47.8 & 5.5 & 16.9 & 47.8 & 7.4 & 17.3 & 27.9 & 50.3 & 86.3 & 144.3 \\
    NGC~1672 & 37.2 & 38.1 & 39.3 & 37.7 & 38.3 & 39.4 & 14.3 & 39.6 & 112.5 & 6.4 & 19.6 & 53.5 & 5.6 & 19.6 & 54.5 & 8.3 & 18.5 & 32.1 & 63.9 & 87.2 & 162.9 \\
    NGC~2835 & 37.0 & 37.7 & 38.5 & 37.4 & 38.0 & 38.7 & 12.1 & 27.2 & 64.5 & 5.7 & 13.0 & 32.2 & 4.2 & 12.6 & 32.0 & 5.3 & 12.4 & 23.8 & 49.1 & 79.8 & 119.8 \\
    NGC~3351 & 37.0 & 37.6 & 39.0 & 37.2 & 37.9 & 39.1 & 9.9 & 19.2 & 76.7 & 4.6 & 10.1 & 36.2 & 3.7 & 9.5 & 36.7 & 4.4 & 8.6 & 13.1 & 38.6 & 65.8 & 104.1 \\
    NGC~3627 & 37.4 & 38.3 & 39.4 & 37.8 & 38.6 & 39.4 & 15.0 & 45.2 & 108.1 & 7.5 & 23.7 & 50.2 & 7.1 & 23.7 & 50.3 & 10.2 & 19.1 & 27.4 & 49.5 & 81.9 & 132.8 \\
    NGC~4254 & 37.4 & 38.2 & 39.0 & 37.7 & 38.4 & 39.0 & 15.2 & 38.7 & 81.5 & 7.3 & 20.3 & 40.9 & 6.2 & 20.0 & 40.7 & 8.6 & 16.4 & 25.7 & 47.5 & 75.2 & 120.5 \\
    NGC~4303 & 37.3 & 38.2 & 39.1 & 37.7 & 38.5 & 39.3 & 14.2 & 38.4 & 100.5 & 6.8 & 18.6 & 47.1 & 5.5 & 18.2 & 47.1 & 8.6 & 16.8 & 28.1 & 48.3 & 84.2 & 145.4 \\
    NGC~4321 & 37.2 & 38.0 & 39.1 & 37.7 & 38.3 & 39.2 & 13.0 & 29.6 & 85.6 & 5.8 & 13.2 & 41.6 & 5.2 & 13.2 & 41.7 & 5.2 & 13.8 & 23.3 & 60.6 & 79.4 & 137.3 \\
    NGC~4535 & 37.2 & 37.8 & 38.7 & 37.5 & 38.1 & 38.8 & 13.1 & 27.3 & 63.2 & 6.3 & 13.9 & 30.1 & 4.9 & 13.4 & 29.8 & 6.2 & 15.3 & 21.2 & 43.4 & 67.7 & 112.6 \\
    NGC~5068 & 37.1 & 37.6 & 38.1 & 37.3 & 37.8 & 38.2 & 11.4 & 24.8 & 41.0 & 5.0 & 12.7 & 20.3 & 4.0 & 12.5 & 20.1 & 5.1 & 11.1 & 18.4 & 36.6 & 49.8 & 73.0 \\
    NGC~7496 & 37.1 & 38.1 & 39.0 & 37.4 & 38.2 & 39.0 & 13.0 & 37.3 & 95.5 & 6.3 & 16.7 & 44.9 & 4.8 & 16.2 & 46.1 & 7.3 & 19.2 & 27.2 & 57.0 & 89.3 & 145.5 \\
    All & 37.2 & 38.0 & 39.1 & 37.6 & 38.3 & 39.1 & 13.3 & 33.1 & 93.8 & 6.2 & 16.4 & 44.2 & 5.1 & 16.3 & 44.4 & 7.1 & 16.6 & 28.3 & 50.3 & 79.4 & 136.7 \\

 \hline
\end{tabular}
\tablefoot{We report percentiles (10\%, 50\%, 90\%) for the H$\alpha$ luminosities measured with HST and MUSE, as well as for several size definitions: circularized radius ($r_{\rm circ}$, HST), moment-based radius ($r_{\rm mom}$, HST), deconvolved moment-based radius ($r_{\rm mom,deconv}$, HST), Stromgren radius ($r_{\rm str}$, HST), and circularized radius from MUSE ($r_{\rm circ,\ MUSE}$). 
Note that the values for the MUSE sample differ from those in \citet{Groves2023}, as here we restrict to the sub-sample also detected with HST in each galaxy (see Tab.\,\ref{tab_map_compprops}).}
\end{sidewaystable*}

\begin{table}
\caption{Distribution of electron densities ($n_{\rm e}$) and ionizing photon production rates ($Q$) for \hii\ regions in each galaxy.}
\label{tab_nQ}
\centering
\begin{tabular}{l ccc ccc}
\hline \hline
Galaxy & \multicolumn{3}{c}{$n_{\rm e}$} & \multicolumn{3}{c}{$Q$} \\

& \multicolumn{3}{c}{[cm$^{-3}$]} & \multicolumn{3}{c}{[log(s$^{-1}$)]} \\
 & 10\% & 50\% & 90\% & 10\% & 50\% & 90\% \\
\hline
\hline

    IC~5332 & 25.0 & 38.8 & 71.9 & 48.7 & 49.3 & 49.9 \\
    NGC~0628 & 19.7 & 38.8 & 78.2 & 48.9 & 49.5 & 50.3 \\
    NGC~1087 & 15.9 & 29.2 & 52.7 & 49.1 & 49.8 & 50.8 \\
    NGC~1300 & 26.7 & 40.5 & 94.6 & 49.1 & 49.6 & 50.5 \\
    NGC~1365 & 25.6 & 42.3 & 163.5 & 49.2 & 50.1 & 51.6 \\
    NGC~1385 & 20.5 & 31.8 & 57.0 & 49.2 & 50.1 & 51.2 \\
    NGC~1433 & 25.1 & 45.2 & 84.8 & 48.9 & 49.5 & 50.4 \\
    NGC~1512 & 22.1 & 57.6 & 98.3 & 48.8 & 49.6 & 50.8 \\
    NGC~1566 & 25.6 & 42.3 & 83.2 & 49.2 & 50.0 & 51.0 \\
    NGC~1672 & 21.5 & 40.5 & 157.5 & 49.1 & 50.0 & 51.2 \\
    NGC~2835 & 19.7 & 31.8 & 67.0 & 48.9 & 49.6 & 50.4 \\
    NGC~3351 & 42.3 & 76.5 & 249.5 & 48.9 & 49.4 & 50.9 \\
    NGC~3627 & 24.0 & 45.2 & 91.1 & 49.4 & 50.2 & 51.2 \\
    NGC~4254 & 24.0 & 40.5 & 85.3 & 49.2 & 50.1 & 50.8 \\
    NGC~4303 & 24.0 & 40.5 & 79.9 & 49.2 & 50.0 & 51.0 \\
    NGC~4321 & 31.8 & 55.1 & 122.1 & 49.1 & 49.9 & 50.9 \\
    NGC~4535 & 27.9 & 50.4 & 103.9 & 49.0 & 49.8 & 50.6 \\
    NGC~5068 & 15.5 & 25.6 & 63.1 & 49.0 & 49.5 & 50.0 \\
    NGC~7496 & 24.0 & 37.1 & 70.1 & 49.0 & 49.9 & 50.8 \\
    All & 22.4 & 40.5 & 87.2 & 49.1 & 49.9 & 50.9 \\

 \hline
\end{tabular}
\tablefoot{We report the 10\%, 50\%, and 90\% percentile values.}
\end{table}

%% file: tables/tab_agemass_assoc.tex
\begin{table}
\caption{Distribution of the age and masses of (NUV, 32\,pc scale) stellar associations across the galaxies within \hii\ regions only (see \S\,\ref{subsec_stellarprops} for details).}
\label{tab_agemass_assoc}
\centering
\begin{tabular}{lcccccc}
\hline\hline
Galaxy & \multicolumn{3}{c}{Age [Myr]} & \multicolumn{3}{c}{Mass [log$_{10}$(\msun)]} \\
       & 10\% & 50\% & 90\% & 10\% & 50\% & 90\% \\
\hline
    IC~5332 & 1.0 & 1.0 & 4.6 & 3.04 & 3.63 & 4.21 \\
    NGC~0628 & 1.0 & 2.0 & 8.4 & 3.20 & 3.86 & 4.61 \\
    NGC~1087 & 1.0 & 2.0 & 8.0 & 3.48 & 4.14 & 4.92 \\
    NGC~1300 & 1.0 & 4.0 & 6.2 & 3.31 & 3.93 & 4.55 \\
    NGC~1365 & 1.0 & 3.0 & 7.5 & 3.62 & 4.37 & 5.36 \\
    NGC~1385 & 1.0 & 4.0 & 8.0 & 3.48 & 4.21 & 5.02 \\
    NGC~1433 & 1.0 & 1.0 & 6.0 & 3.08 & 4.04 & 4.58 \\
    NGC~1512 & 1.0 & 1.0 & 7.6 & 3.43 & 4.05 & 4.82 \\
    NGC~1566 & 1.0 & 2.0 & 7.0 & 3.63 & 4.19 & 5.09 \\
    NGC~1672 & 1.0 & 3.0 & 5.0 & 3.52 & 4.31 & 5.15 \\
    NGC~2835 & 1.0 & 2.0 & 8.0 & 3.18 & 3.97 & 4.68 \\
    NGC~3351 & 1.0 & 3.0 & 7.0 & 3.27 & 4.01 & 5.44 \\
    NGC~3627 & 1.0 & 4.0 & 7.9 & 3.70 & 4.42 & 5.21 \\
    NGC~4254 & 1.0 & 4.0 & 7.0 & 3.59 & 4.34 & 5.11 \\
    NGC~4303 & 1.0 & 3.0 & 7.0 & 3.58 & 4.32 & 5.16 \\
    NGC~4321 & 1.0 & 3.0 & 7.0 & 3.59 & 4.29 & 5.14 \\
    NGC~4535 & 1.0 & 4.0 & 8.0 & 3.53 & 4.18 & 5.04 \\
    NGC~5068 & 1.0 & 7.0 & 10.0 & 3.10 & 3.81 & 4.68 \\
    NGC~7496 & 1.0 & 1.0 & 5.2 & 3.52 & 4.04 & 4.88 \\
    All & 1.0 & 3.0 & 7.0 & 3.49 & 4.22 & 5.05 \\
\hline
\end{tabular}
\tablefoot{We report the 10\%, 50\%, and 90\% percentile values.}
\end{table}

%% file: figures/fig_noise.tex
\begin{figure*}
    \centering
        \includegraphics[trim={0 1.2cm 0 0},clip, width=\textwidth]{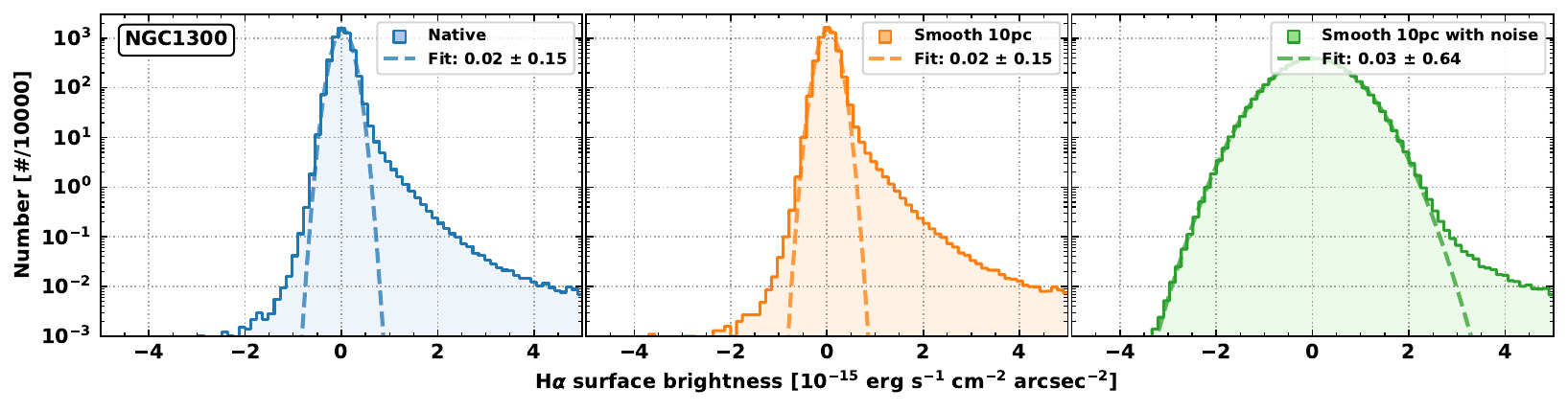} \\ \vspace{-2mm}
        \includegraphics[width=\textwidth]{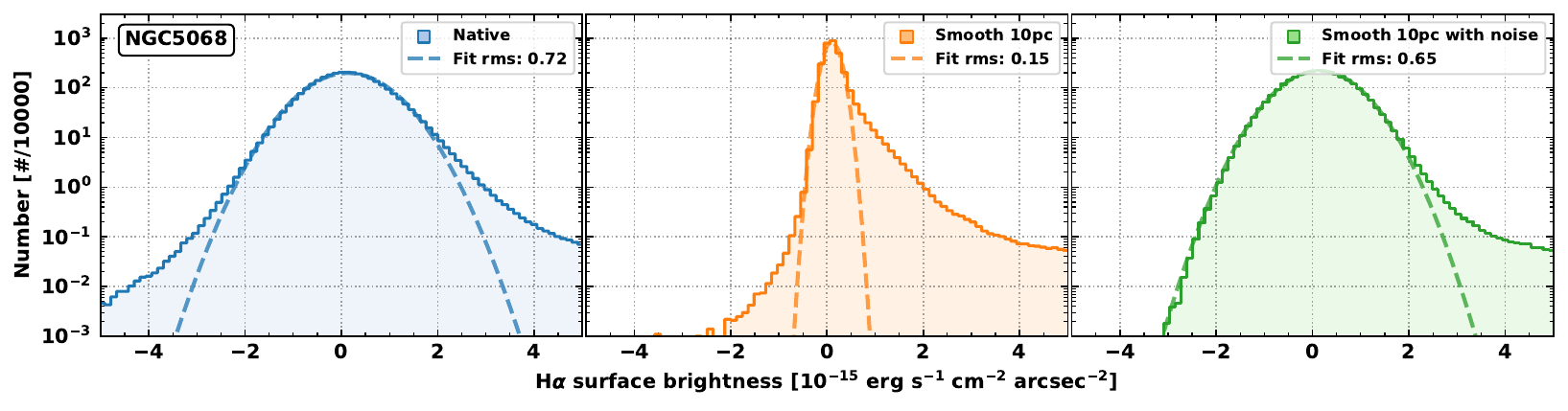} \vspace{-5mm}
    \caption{\textbf{Pixel-value distributions of H$\alpha$ surface brightness for two representative galaxies.} 
Shown are histograms of the HST H$\alpha$ images for NGC\,1300 (upper panels) and NGC\,5068 (lower panels), comparing the native data (left), the data convolved to a fixed physical resolution of 10\,pc (middle), and the convolved data with additional noise added (right). 
Histograms are plotted as step lines with shaded areas, and Gaussian fits to the pixel distributions are overlaid as dashed curves (the fit standard deviation - approximately equivalent to the rms noise level of the image - is given in the legend of each panel in units of $10^{-15}$ erg s$^{-1}$ cm$^{-2}$ arcsec$^{-2}$.}
    \label{fig_noise}
\end{figure*}